\documentclass[aps,prd,nofootinbib,twocolumn,superscriptaddress,preprintnumbers]{revtex4-1}

\usepackage{amssymb}
\usepackage{amsmath}
\usepackage{slashed}
\usepackage{hyperref}
\usepackage{breakurl}
\usepackage{bm}
\usepackage{color}
\usepackage{tikz}
\usepackage[utf8]{inputenc}
\usetikzlibrary{snakes}
\usetikzlibrary{decorations}
\usetikzlibrary{trees}
\usetikzlibrary{decorations.pathmorphing}
\usetikzlibrary{decorations.markings}
\usetikzlibrary{external}
\usetikzlibrary{intersections}
\usetikzlibrary{shapes,arrows}
\usetikzlibrary{arrows.meta}
\usetikzlibrary{shapes.misc}
\usetikzlibrary{decorations.text}
\usetikzlibrary{backgrounds}
\usetikzlibrary{fadings}
\usetikzlibrary{tikzmark,calc,arrows,shapes,decorations.pathreplacing}

\usepackage{subcaption}
\usepackage{tabstackengine}
\stackMath

\makeatletter
\long\def\@makecaption#1#2{%
  \par
  \vskip\abovecaptionskip
  \begingroup
    \small\rmfamily
    \sbox\@tempboxa{%
      \let\\\heading@cr
      \@make@capt@title{#1}{#2}%
    }%
    \@ifdim{\wd\@tempboxa >\hsize}{%
      \begingroup
        \samepage
        \flushing
        \let\footnote\@footnotemark@gobble
        \@make@capt@title{#1}{#2}\par
      \endgroup
    }{%
      \hb@xt@\hsize{\unhbox\@tempboxa\hfil}%
    }%
  \endgroup
  \vskip\belowcaptionskip
}%
\makeatother

\makeatletter
\def\simgt{\mathrel{\lower2.5pt\vbox{\lineskip=0pt\baselineskip=0pt
           \hbox{$>$}\hbox{$\sim$}}}}
\def\simlt{\mathrel{\lower2.5pt\vbox{\lineskip=0pt\baselineskip=0pt
           \hbox{$<$}\hbox{$\sim$}}}}

\def\fig#1{Fig.~\ref{#1}}

\newcommand{\figsubref}[2]{%
  \hyperref[#2]{\ref*{#1}\subref*{#2}}%
}
\makeatother

\def\eqn#1{Eq.~\eqref{#1}}

\def\Section#1{\vskip .15 cm 
\noindent {\it #1---}}
\newcommand{\be}{\begin{equation}}
\newcommand{\ee}{\end{equation}}

\def\deltabar{{\mathchar '26\mkern -10mu\delta}}

\renewcommand{\imath}{\mathrm{i}}

\def\topbotatom#1{\hbox{\hbox to 0pt{$#1\bot$\hss}$#1\top$}}

\allowdisplaybreaks

\begin{document}

\title{Classical Gravitational Scattering from the Ultraviolet and\\[2pt]
the Absence of Calabi--Yau Integrals in the Conservative Sector at $\mathcal O(G^5)$}

\author{Zvi Bern}
\affiliation{
Mani L. Bhaumik Institute for Theoretical Physics,
University of California at Los Angeles,
Los Angeles, CA 90095, USA}

\author{Avery Jackman} 
\affiliation{
Mani L. Bhaumik Institute for Theoretical Physics,
University of California at Los Angeles,
Los Angeles, CA 90095, USA}

\author{Gareth Mansfield} 
\affiliation{
Mani L. Bhaumik Institute for Theoretical Physics,
University of California at Los Angeles,
Los Angeles, CA 90095, USA}

\author{Michael~S.~Ruf}
\affiliation{
SLAC National Accelerator Laboratory, Stanford University, Stanford, CA 94309, USA
}

\begin{abstract}
We explain why Calabi--Yau and complete elliptic integrals do not contribute to conservative observables at fifth post-Minkowskian order, despite appearing in intermediate steps. At even loop orders, conservative contributions are tied to terms proportional to the logarithm of the momentum transfer, which in dimensional regularization arise from singular regions. We show that in the classical limit, the integral classes responsible for Calabi--Yau and complete elliptic behavior are absent from the ultraviolet singular structures that generate the required logarithm. This perspective also suggests alternative strategies for analyzing the classical limit of multiloop integrals in the conservative sector at even loop orders.

\end{abstract}

\maketitle

\Section{Introduction}
\label{sec:intro}
The expected dramatic gains in sensitivity and frequency coverage for planned gravitational-wave detectors~\cite{Punturo:2010zz, LISA:2024hlh, Reitze:2019iox, LIGOasharp, Abac:2025saz} demand high-precision theoretical modeling across a large parameter space.   Meeting this challenge will require significant advances in multiple complementary theoretical frameworks, including numerical relativity~\cite{Pretorius:2005gq, Campanelli:2005dd, Baker:2005vv, Damour:2014afa},  the gravitational self-force (SF) program~\cite{Mino:1996nk, Quinn:1996am, Poisson:2011nh, Barack:2018yvs}, effective field theory (EFT)~\cite{Goldberger:2004jt, Cheung:2018wkq}, and both post-Newtonian (PN)~\cite{Droste:1916, Droste:1917, Einstein:1938yz, Ohta:1973je, Blanchet:2013haa} and post-Minkowskian (PM)~\cite{Bertotti:1956pxu, Kerr:1959zlt, Bertotti:1960wuq, Westpfahl:1979gu, Portilla:1980uz, Bel:1981be} methods. The PM approach, which is the framework of the present Letter, has seen rapid progress in recent years,  with computations at third (3PM)~\cite{Bern:2019nnu, Bern:2019crd}, fourth (4PM)~\cite{Bern:2021dqo, Bern:2021yeh, Dlapa:2021npj, Dlapa:2021vgp, Damgaard:2023ttc}, and more recently, new computations have been pushing the frontier to the fifth order (5PM) in the expansion~\cite{Bern:2023ccb, Bern:2024adl, 
Driesse:2024xad, Driesse:2024feo, Bini:2025vuk, Bern:2025zno, Bern:2025wyd, Driesse:2026qiz}.   

Here we explain a striking feature encountered in the conservative part of the 5PM calculations.  Higher-loop orders typically introduce increasingly complicated classes of special functions. For example, at 3PM one encounters logarithms; at 4PM, in addition to generalized polylogarithms (GPLs)~\cite{Goncharov:2001iea}, symmetric-square K3 complete elliptic integrals appear~\cite{Bern:2021dqo}. By 5PM, both complete elliptic and Calabi--Yau (CY) three-fold integrals appear in intermediate steps~\cite{Frellesvig:2023bbf, Klemm:2024wtd}, yet once the various contributions are combined, they cancel in conservative parts. This pattern is observed in both maximal supergravity and Einstein gravity~\cite{Bern:2024adl, Driesse:2024xad, Driesse:2024feo, Bern:2025zno, Bern:2025wyd, Driesse:2026qiz}. 

We show that this cancellation is not accidental. By reorganizing the calculation, the absence of the complete elliptic and CY contributions becomes manifest.  In the context of the amplitudes-based approach, one obtains a classical integrand by first expanding in soft graviton momenta~\cite{Cheung:2018wkq, Bern:2019nnu, Bern:2019crd}.
Then standard multiloop integration-by-parts (IBP) methods are  applied~\cite{Chetyrkin:1981qh, Tkachov:1981wb, Laporta:2000dsw}, after which the large number of loop integrals is reduced to a much smaller basis of master integrals~\cite{Henn:2013pwa, Smirnov:2020quc, Usovitsch:2020jrk}. This reduction procedure is computationally demanding and typically requires specialized implementations (see, e.g., Refs.~\cite{Smirnov:2008iw, Maierhofer:2017gsa, Guan:2024byi, FIRE7, Lange:2025fba, Lee:2013mka}). The master integrals are then evaluated using differential equations~\cite{Kotikov:1990kg, Bern:1993kr, Remiddi:1997ny, Gehrmann:1999as, Henn:2013pwa}.  While this framework is extremely powerful, it can obscure underlying structural features, specifically, why the most complicated integral functions cancel out of the final expressions.

At even loop orders, within the soft graviton region relevant for classical physics, conservative contributions are necessarily proportional to a logarithm of the momentum transfer. Such logarithms only arise when ultraviolet or infrared divergences are present in the soft region (see discussion below). By focusing on the ultraviolet divergences in this region, one can directly expose the absence of complete elliptic and CY integrals in the conservative part of 5PM.  

More generally, ultraviolet (see, e.g., Refs.~\cite{Caron-Huot:2016cwu, Herzog:2017ohr, Bern:2018jmv, Bern:2020ikv}) and infrared divergences (see, e.g., Refs.~\cite{Weinberg:1965nx, Naculich:2011ry, Akhoury:2011kq, Anastasiou:2018rib, Anastasiou:2020sdt, Heissenberg:2021tzo})  are typically simpler to evaluate than finite parts. Although complete elliptic and Calabi--Yau integrals need not cancel in finite regions, and indeed they do not, we show that at 5PM order they are entirely absent in the singular regions governing conservative classical dynamics. The relative simplicity of these regions, therefore, suggests a more efficient route to computing complete conservative contributions at even loop orders. Although not the subject of the present paper, we verified that assembling the ultraviolet and infrared singularities reproduces the known 3PM conservative result of Ref.~\cite{Bern:2019nnu}.  Similar considerations should be applicable to extract the energy loss (see e.g., Refs.~\cite{Damour:2020tta, Herrmann:2021tct, Dlapa:2022lmu, Driesse:2024feo}) at odd-loop orders.

\begin{figure}[tb]
\begin{subfigure}{0.2\columnwidth}
  \centering
  \includegraphics[width=.9 \linewidth]{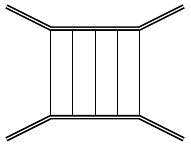}
  \caption{}
  \label{subfig:ladder}
\end{subfigure}
\begin{subfigure}{0.2\columnwidth}
\centering
  \includegraphics[width=.9\linewidth]{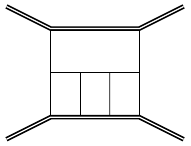}
    \caption{}
    \label{subfig:Diag391}
\end{subfigure}
\begin{subfigure}{0.2\columnwidth}
\centering
  \includegraphics[width=.9\linewidth]{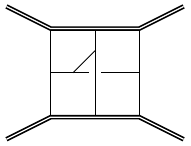}
    \caption{}
    \label{subfig:Diag360}
\end{subfigure}
\begin{subfigure}{0.2\columnwidth}
\centering
  \includegraphics[width=.9\linewidth]{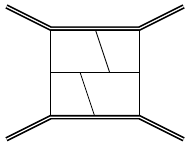}
    \caption{}
    \label{subfig:Diag366}
\end{subfigure}
\vskip -.2 cm 
  \caption{Examples of parent diagrams with only cubic vertices relevant for classical scattering at $\mathcal{O}(G^5)$.  The double lines represent scalar matter, and the single lines represent gravitons.
  }
   \label{fig:ParentIntegrals}
\end{figure}

\Section{Classical gravitational amplitudes}
In the amplitudes-based approach, which we rely on in this Letter, the relativistic $2\!\to\!2$ scattering amplitudes of scalar particles are used to represent the scattering of two nonspinning black holes or other compact objects. Because the classical limit of the scattering amplitude is connected to the radial action~\cite{Bern:2021dqo}, the amplitude is directly related to the conservative scattering angle. Here, we are concerned with scattering processes at even loop orders corresponding to odd orders in the PM expansion. The structure of such gravitational amplitudes has been extensively analyzed at two and four loops; for further details, see, for example, Refs.~\cite{Bern:2019crd, Bern:2021dqo, Bern:2021yeh, Bern:2025wyd}.   A variety of methods exist for connecting amplitudes to physical observables~\cite{Cheung:2018wkq, Kosower:2018adc, Bern:2019nnu,  Bern:2019crd, Bern:2021dqo, Damgaard:2023ttc, Brandhuber:2025igz, Kim:2025gis}. 

Loop integrands are constructed using generalized unitarity~\cite{Bern:1994zx, Bern:1994cg, Bern:1997sc, Britto:2004nc, Bern:2004cz, Bern:2007ct} together with the double-copy framework~\cite{Kawai:1985xq, Bern:2008qj, Bern:2010ue, Bern:2019prr}, starting from tree-level gauge-theory amplitudes. The classical limit is obtained by expanding the integrand in the regime of small momentum transfer, $q \ll m_i$, and systematically separating potential and radiation regions using the method of regions~\cite{Beneke:1997zp}. Prior to integration, the integrand can be simplified further by mapping it onto an integrand basis~\cite{Bern:2024vqs}.  

The resulting expressions are usually best evaluated using techniques based on integration by parts~\cite{Chetyrkin:1981qh, Tkachov:1981wb} and differential equations~\cite{Kotikov:1990kg, Bern:1993kr, Remiddi:1997ny, Gehrmann:1999as, Henn:2013pwa}.  It is advantageous to employ kinematic variables tailored to the classical limit of relativistic scattering amplitudes~\cite{Parra-Martinez:2020dzs}, as we do here. The process is parametrized by the four-velocities $u_i$, masses $m_i$ ($i=1,2$), and the momentum transfer $q$, subject to the constraint $u_i \!\cdot q = 0$. In this parametrization, the relativistic loop integrals depend only on the invariant $\sigma = u_1 \!\cdot u_2$.

\Section{Integral types and diagram topologies}
Multiloop calculations are typically organized into families of integrals, defined by parent-diagram topologies with the maximal number of propagators. At 5PM order, these parent diagrams contain 13 propagators.  We will refer to the remaining integral topologies that follow from the parent diagrams by collapsing propagator lines as daughter diagrams.  Representative examples from the four-loop radial-action computation of Ref.~\cite{Bern:2025wyd} are displayed in Fig.~\ref{fig:ParentIntegrals}. In classifying the relevant integral families, it suffices to consider planar topologies, with a single exception corresponding to the nonplanar parent diagram (d) whose nonplanarity does not involve the matter lines. This simplification arises because nonplanar contributions recombine with planar ones into eikonal sums, effectively replacing certain matter propagators with on-shell delta functions (see below and Refs.~\cite{Saotome:2012vy, Akhoury:2013yua, Bern:2024adl, Bern:2025zno, Bern:2025wyd}). Consequently, in the discussion below, one matter propagator per loop is taken to be on shell, a feature that is manifest in worldline formalisms~\cite{Goldberger:2004jt, Jakobsen:2021zvh, Driesse:2024xad}.

\begin{figure}[tb]

\centering
\begin{subfigure}{0.19\columnwidth}
\setcounter{subfigure}{4}
  \centering
  \includegraphics[width=0.9\linewidth]{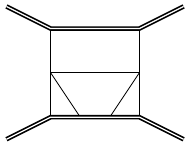}
  \caption{}
   \label{subfig:Ell}
\end{subfigure}\hfill
\begin{subfigure}{0.19\columnwidth}
  \centering
  \includegraphics[width=0.9\linewidth]{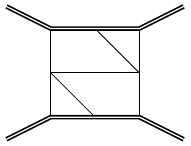}
  \caption{}
  \label{subfig:CY}
\end{subfigure}\hfill
\begin{subfigure}{0.19\columnwidth}
  \centering
  \includegraphics[width=0.9\linewidth]{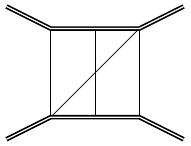}
  \caption{}
   \label{subfig:Heun}
\end{subfigure}\hfill
\begin{subfigure}{0.19\columnwidth}
  \centering
  \includegraphics[width=0.9\linewidth]{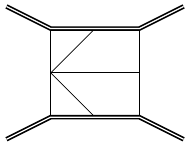}
  \caption{}
   \label{subfig:Polylog}
\end{subfigure}\hfill
\begin{subfigure}{0.19\columnwidth}
  \centering
  \includegraphics[width=0.9\linewidth]{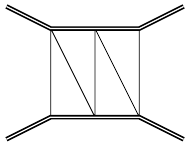}
  \caption{}
   \label{subfig:ZigZag}
\end{subfigure}\hfill
\vskip -.2 cm 
\caption{Representative $\mathcal{O}(G^5)$ classical-scattering integral topologies. Diagrams \subref{subfig:Ell}--\subref{subfig:Heun} generate functions beyond GPLs: \subref{subfig:Ell} yields elliptic, \subref{subfig:CY} CY, and \subref{subfig:Heun} Heun integrals. In contrast, \subref{subfig:Polylog} and \subref{subfig:ZigZag} evaluate to GPLs.
}
\label{fig:integrals}
\end{figure}

The classes of special functions that may appear in an amplitude are directly tied to the topology of the diagrams defining the loop integrals (see e.g., Refs.~\cite{Duhr:2019tlz, Weinzierl:2022eaz, Bourjaily:2022bwx}). These functions can, at least formally, be expressed in terms of Chen iterated integrals~\cite{Chen:1977oja},
\begin{equation}
\Gamma_{i_1,\ldots,i_n}(x) = \int_1^{x} \mathrm{d}x_n\, k_{i_n}(x_n) \cdots
\int_1^{x_2} \mathrm{d}x_1\, k_{i_1}(x_1)\,,
\end{equation}
for a set of integration kernels $k_i(x)$, normalized by period integrals related to the leading singularities of the integrals. The analytic structure of these functions, including their numerical evaluation and analytic continuation, is dictated by the nature of the kernels.
In the remainder of this letter, we will discuss three classes of kernels, which are determined by the maximal cut of the integrals depicted in \fig{fig:integrals} :
\begin{itemize}
\item ``Elliptic"  \ref{fig:integrals}\subref{subfig:Ell}: The maximal cut involves an integration over a K3 surface, which evaluates to the square of a complete elliptic integral of the first kind \cite{Brammer:2025rqo},
\item ``Heun" \ref{fig:integrals}\subref{subfig:Heun}: The maximal cut gives rise to a distinct K3 surface where the integral evaluates to a product of Heun\footnote{We define these as functions related to a geometry whose period satisfies a second-order Heun differential equation. Notably, these do not cancel from the final result, distinguishing them from simpler elliptic cases.} functions~\cite{MRAmplitudes2023, Brammer:2025rqo, GSJoyce_2004, Ronveaux:1995Heun, Ablinger:2017bjx},
\item ``Calabi-Yau" \ref{fig:integrals}\subref{subfig:CY}: This involves integration over a CY three-fold geometry \cite{Frellesvig:2023bbf, Klemm:2024wtd}.
\end{itemize}
In contrast, the maximal cuts of other topologies, such as those in classes \figsubref{fig:integrals}{subfig:Polylog} and \figsubref{fig:integrals}{subfig:ZigZag}, are purely algebraic. Such integrals evaluate to generalized polylogarithms (GPLs) \cite{Goncharov:2001iea} to all orders in the dimensional regulator parameter $\epsilon = (4-D)/2$. Furthermore, any diagram that is not a parent of the topologies shown in \figsubref{fig:integrals}{subfig:Ell}--\subref{subfig:Heun} generates only GPLs. For instance, the parent ladder diagram in \fig{fig:ParentIntegrals}\subref{subfig:ladder} and its descendants are restricted to the GPL class and cannot contribute to more complex function spaces, a result confirmed by explicit calculation \cite{Bern:2023ccb}.

We will not need detailed properties of these functions beyond their link to specific diagram topologies in \fig{fig:integrals}. By analyzing UV properties, we can deduce the absence of certain special functions. For this purpose, it is better to directly focus on the contact topologies in Fig.~\figsubref{fig:integrals}{subfig:Ell}--\figsubref{fig:integrals}{subfig:Heun}, rather than the parent diagrams in \fig{fig:ParentIntegrals}. While parent diagrams can contribute to CY/elliptic functions, they do so through inhomogeneous terms in differential equations sourced by daughter integrals with topologies \figsubref{fig:integrals}{subfig:Ell}--\figsubref{fig:integrals}{subfig:Heun}. The homogeneous system is defined by maximal cuts, i.e., by setting all daughter integrals to zero. At four loops, for all parent integrals, these homogeneous systems are solvable within the space of GPLs, with no appearance of the more complicated integrals~\cite{Bern:2023ccb, Bern:2024adl, Driesse:2024xad, Bern:2025zno, Bern:2025wyd}. Thus, focusing on the contact topologies in \fig{fig:integrals} is sufficient to determine whether CY/elliptic special functions appear in the classical amplitude.

\Section{Classical limits and divergences at even loop orders}
At even loop order $L$ (corresponding to odd powers of Newton's constant $G$), the classical conservative contribution to the amplitude, $\mathcal{M}_{\rm cl}$, is associated with a logarithmic branch cut, which is absent at odd loop orders. Schematically,
\begin{equation}
\begin{aligned}
\mathcal{M}_{\rm cl}
\sim
\begin{cases}
(-q^2)^{\tfrac{L}{2}-1}\,\ln(-q^2)\,, & L \ \text{even}\,,\\[4pt]
(-q^2)^{\tfrac{L}{2}-1}\,,           & L \ \text{odd}\,.
\end{cases}
\end{aligned}
\label{eq:ClassicalTerms}
\end{equation}
This structure follows directly from the position-space version: the $L$-loop term in the post-Minkowskian expansion generates a potential proportional to $(G/r)^{L+1}$, whose Fourier transform produces precisely the behavior in Eq.~\eqref{eq:ClassicalTerms}. The power counting in Eq.~\eqref{eq:ClassicalTerms} depends on the observable under consideration; for example, even and odd loop orders are interchanged for the radiated momentum. Similar modifications arise when including spin effects or working in different spacetime dimensions.

A central observation is that, in dimensional regularization, the required $\ln(-q^2)$ can arise only in conjunction with contributions singular in the dimensional-regularization parameter $\epsilon$. Dimensional analysis fixes the overall $q^2$ dependence of the amplitude, with the logarithm emerging from the expansion of singular terms,
\begin{equation}
\frac{1}{\epsilon}\,(-q^2)^{-L\epsilon}
= \left[\frac{1}{\epsilon} - L\ln(-q^2) + \mathcal{O}(\epsilon)\right]\,.
\label{eq:LogDivRelation}
\end{equation}
The $1/\epsilon$ pole itself corresponds to a local contribution and can be discarded when focusing on long-distance physics;  nevertheless, it tags the desired $\ln(-q^2)$. As a result, the logarithmic term of interest is entirely determined by the singular part of the amplitude. In particular, at four loops, the coefficient of $(-q^2) \ln(-q^2)$ can be obtained by computing the coefficient of the $(-q^2)/\epsilon$ divergence.
Similar ideas for extracting logarithms from singularities have been used in QCD in the context of collider physics (see, e.g., Refs.\cite{Mitov:2006xs, Baikov:2016tgj}).

The observation that the classical amplitude is entirely determined by divergent contributions immediately suggests an alternative strategy for obtaining the relevant classical terms.  In what follows, we exploit this connection by focusing on the divergent terms in the amplitude, which suffice to explain the absence of certain classes of special functions in classical observables. 


\Section{The structure of divergences}
In general, scattering amplitudes contain both infrared (IR) and ultraviolet (UV) divergences, which, despite their distinct physical origins, both appear as poles in the dimensional regulator $\epsilon$. To extract the complete classical contribution to the amplitude at even loop orders, it is therefore necessary, in general, to determine both the UV and IR poles in $\epsilon$. Our analysis is performed prior to applying integration-by-parts identities, which can mix contributions from the UV and IR regions.

Gravity is significantly better behaved in the infrared than gauge theory, owing to the additional powers of loop momenta appearing in the numerators of gravitational Feynman diagrams.   While there are uncanceled IR singularities in classical amplitudes, these arise from iteration terms (see e.g., Ref.~\cite{Cheung:2018wkq, Bern:2019crd}), which are associated with diagrams such as the ladder in \figsubref{fig:ParentIntegrals}{subfig:ladder}, which, when two matter lines are cut, split the diagram into two subdiagrams. Because the integrals arising from such iterations involve only GPLs, we restrict our attention to genuinely classical integrals, excluding iteration contributions, such as those shown in Figs.~\ref{fig:ParentIntegrals}\subref{subfig:Diag391}–\subref{subfig:Diag366}. These integrals may be analyzed entirely in terms of their ultraviolet behavior. Depending on the problem, the integrals may exhibit power-law divergences that vanish in dimensional regularization but require expanding the integrals to higher orders in the UV expansion.   This includes the elliptic and CY sectors of interest, for which a UV analysis is sufficient.

An important consideration is that even for IR-finite diagrams, individual terms within a given diagram can be IR divergent, with the divergences canceling when combined.   For example, interpreting the diagrams in Fig.~\ref{fig:ParentIntegrals}\subref{subfig:Diag391}--\subref{subfig:Diag366} as Feynman diagrams in de Donder gauge results in IR finite expressions for each~\cite{Akhoury:2011kq}. However, each term in the diagram may exhibit IR divergences. Moreover, when numerator factors cancel propagators, contributions to such terms can be reassigned to different diagram topologies. As a result, topologies that are not expected to contain IR singularities may acquire them, only to cancel against corresponding singularities elsewhere.  In what follows, we assume that the diagrams have not been reorganized in a way that introduces spurious IR singularities that cancel, allowing us to perform an analysis based solely on UV behavior.   

Another subtlety is that applying the method of regions~\cite{Beneke:1997zp} to separate potential and radiation modes introduces additional ultraviolet and infrared divergences that cancel only after summing over all regions. For simplicity, we do not perform this separation here, and our results therefore apply directly to the full amplitude in the classical limit. The decomposition into distinct regions is normally used to improve the efficiency of integral evaluations (see, e.g., Ref.~\cite{Bern:2019crd}). Implementing this refinement is straightforward and could further streamline the extraction of complete classical amplitudes; we leave this direction for future work.

In general, integrals may contain subdivergences that do not cancel even when all contributions are combined. When present, such subdivergences are associated with finite-size effects, as the corresponding divergences map onto higher-dimensional operators that encode such physics. These effects are captured by local counterterms, whose inclusion is straightforward in the analysis presented here and will not appear at four loops, due to the too high dimension of the associated tidal operators (see e.g., comments in Refs.~\cite{Barack:2023oqp, Ivanov:2024sds, Caron-Huot:2025tlq}).


\Section{Integrals at $\mathcal{O}(G^3)$ from UV expansion}
We illustrate the method with a two-loop example. This example is chosen for simplicity, to elucidate the common features of integrals that arise in physical problems, such as the scattering angle in general relativity.
To be concrete, consider the integral,
\begin{equation}
\vcenter{\hbox{\includegraphics[scale=.4]{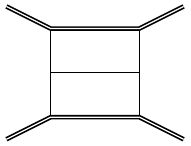}}}\!\!= \!\int\!\frac{\mathrm{d}^D\ell_1\mathrm{d}^D\ell_2}{(\imath\pi^{D/2})^2}\frac{\deltabar(2u_1\cdot \ell_1)\, \deltabar(2u_2\cdot \ell_2)\, \mathcal{N}}{\ell_1^2\ell_2^2(\ell_1{-}q)^2(\ell_2{-}q)^2(\ell_1{+}\ell_2{-}q)^2}\,.
\label{eq:Hintegral}
\end{equation}
Here and in the following, double lines will represent cut eikonal propagators of the form $\deltabar(2u_1\cdot \ell_i) \equiv 2\pi \imath\delta(2u_1\cdot \ell_i)$ for the top and  $\deltabar(2u_2\cdot \ell_i)$ for the bottom line, while thin lines are quadratic massless propagators.
In the diagrams, we implicitly assume the presence of the numerator $\mathcal{N}$, which renders the integral IR finite, as happens for the four-loop integrals of interest. For the example at hand, to be explicit, we take the numerator to be 
\begin{equation}
\mathcal{N}=\ell_1\cdot(\ell_1-q)\times\ell_2\cdot(\ell_2-q)\,.
\end{equation}
We take this as a simpler stand-in for the numerator that arises for the given gravity Feynman diagram.
With the chosen numerator, or equivalently, for the one that arises in gravity, the integral does not contain an infrared singularity. Thus, the $1/\epsilon$ pole of the integral can be determined uniquely from the UV divergent part of the integrand. In this limit, $|q|\ll |\ell_i|$, greatly simplifying its evaluation. 

However, if one proceeds straightforwardly, a technical problem arises with dimensional regularization: in the limit, the integrals become scaleless, implying that the integral vanishes. This occurs because the integral develops an infrared divergence that cancels the ultraviolet one.  To separate the desired ultraviolet divergences from the spurious infrared ones, we reintroduce a scale that serves as an infrared cutoff without affecting the UV one, and compute the resulting divergence.  We do so by shifting the arguments of the delta functions by a small mass parameter $\Delta$. This can be interpreted as uniformly pushing the external matter lines off shell in the amplitude, which then enables the extraction of the UV divergence via dimensional regularization. This procedure resembles the IR regularization used in Ref.~\cite{Grozin:2015kna}.

Then after expanding in the UV limit while keeping $\ell_i\sim \Delta$, we obtain,
\begin{align}
&\left.\vcenter{\hbox{\includegraphics[scale=.45]{figs/HIntegral1.pdf}}}\right|_{\mathrm{UV}}\label{eq:UVH1}\\[4pt]
={}&\left.\int\frac{\mathrm{d}^D\ell_1\mathrm{d}^D\ell_2} {(\imath\pi^{D/2})^2}\, \frac{\deltabar(2u_1\cdot \ell_1{+}\Delta)\, \deltabar(2u_2\cdot \ell_2{+}\Delta)\, \mathcal{N}}{[\ell_1^2]^{2}[\ell_2^2]^{2}(\ell_1{+}\ell_2)^2}\right|_{\mathrm{UV}}.\nonumber
\end{align}
This equation admits a natural pictorial representation
\begin{equation}
\left.\vcenter{\hbox{\includegraphics[scale=.5]{figs/HIntegral1.pdf}}}\right|_{\mathrm{UV}}=\left.\vcenter{\hbox{\includegraphics[scale=.38]{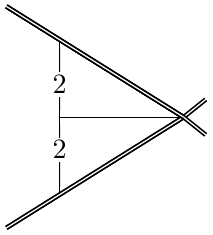}}}\right|_{\mathrm{UV}}\,.\label{eq:UVRelation2Loop}
\end{equation}
Here, a line with a label `$a$' represents a propagator raised to the power, e.g., $(\ell^2)^{-a}$.   For our case, the numerator becomes $\mathcal N \rightarrow \ell_1^2 \ell_2^2$, which cancels one factor from the squared propagator.  In our notation, the diagrams here represent the propagators, keeping the numerators implicit. We observe that the integral simplifies drastically: it is independent of the momentum transfer $q$, and the denominator is correspondingly simplified. 

The integrals on the right-hand side of Eq.~\eqref{eq:UVH1} can be computed using standard techniques for dimensionally regulated integrals, including IBP-reduction and canonical differential equations.  We summarize the computation in the Supplementary Material~\ref{CuspyIntegrals}, with the result, 
\begin{equation}\label{eq:UVIntegral}
\left.\vcenter{\hbox{\includegraphics[scale=.4]{figs/Cuspy4.pdf}}}\right|_{\mathrm{UV}}=\frac {\imath\pi }{\epsilon}\frac{ {\rm arccosh}(\sigma)(\imath\pi-{\rm arccosh}(\sigma))}{2\sqrt{\sigma^2-1}}\,.
\end{equation}
We have also evaluated the complete integral in Eq.~\eqref{eq:Hintegral} using IBP reduction and the results of Refs.~\cite{Herrmann:2021lqe, Herrmann:2021tct} and find agreement for terms proportional to the required pole in $\epsilon$.

Using this example, we have illustrated how a classical integral can be reconstructed from its UV behavior. Analogous considerations apply to IR divergences. We have checked that once IR and UV divergences are taken into account in each class of integrals, the complete $\mathcal{O}(G^3)$ conservative amplitude is reproduced.
Thus, this framework provides an alternative method to compute the conservative scattering angle at $\mathcal{O}(G^3)$, first reported in Refs.~\cite{Bern:2019nnu, Bern:2019crd}.  Here, we limit our analysis to understanding the cancellation of certain integrals.


\Section{Absence of complete elliptic integrals}
The cancellation of elliptic integrals was observed in the 1SF and 2SF contributions to the $\mathcal O(G^5)$ conservative scattering angle in Refs.~\cite{Bern:2024adl, Driesse:2024xad, Bern:2025wyd, Driesse:2026qiz}. We will show that this seemingly surprising cancellation can, in fact, be seen without performing an explicit evaluation of the integrals.  To do so, consider the following integral, which generates the elliptic contribution,
\begin{align}\label{eq:elliptic_diagram}
\mathcal{I}_{\mathrm{ell}}=\vcenter{\hbox{\includegraphics[scale=.5]{figs/EllipticIntegral1.pdf}}}\,.
\end{align}
As for the two-loop example, the precise form of the numerator is unimportant to the discussion, only that it be infrared finite; a worked-out example with an explicit numerator is given in the Supplementary Material~\ref{sec:Subloops}. The triangle subloops can be integrated out, resulting in a two-loop integral with propagators raised to non-integer indices\footnote{Note that there are no sub-UV divergences for one-loop eikonal triangles in dimensional regularization.}
\begin{align}\label{eq:Iell}
\mathcal{I}_{\mathrm{ell}}=\vcenter{\hbox{\includegraphics[scale=.6]{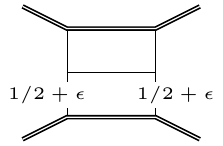}}},
\end{align}
where we have suppressed the factor resulting from the triangle integrations.
Here on the right-hand side, the numerator has changed due to the subloop integration.

Proceeding in the same manner as for the two-loop integral in Eq.~\eqref{eq:UVRelation2Loop}, we obtain
\begin{equation}
\left.{\mathcal{I}_{\mathrm{ell}}}\right|_{\mathrm{UV}}=\left.\vcenter{\hbox{\includegraphics[scale=.4]{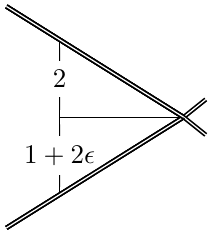}}}\right|_{\mathrm{UV}} \!\!.
\end{equation}
The right-hand side evaluates to multiple polylogarithms. Therefore, the divergent part of $\mathcal{I}_\ell$ does not contain any elliptic functions, as can be seen after setting $\epsilon \rightarrow 0$. The resulting two-loop integrals have been computed in Ref.~\cite{Grozin:2015kna}.


\Section{Absence of Calabi--Yau integrals}
We next consider integrals associated with Calabi--Yau (CY) geometries that could, in principle, contribute to the 5PM amplitude and scattering angle. Such contributions arise from the integral topology~\cite{Frellesvig:2024rea, Klemm:2024wtd},
\begin{equation}
\mathcal{I}_{\mathrm{CY}}
=\vcenter{\hbox{\includegraphics[scale=.55]{figs/CYIntegral1.pdf}}}
=\vcenter{\hbox{\includegraphics[scale=.55]{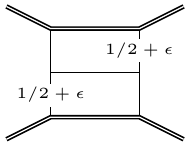}}}\,.
\end{equation}
In this representation, the ultraviolet (UV) expansion is straightforward and after integrating out the triangles yields,
\begin{equation}
\left.
\vcenter{\hbox{\includegraphics[scale=.55]{figs/CYIntegral1.pdf}}}
\right|_{\mathrm{UV}}
=
\left.
\vcenter{\hbox{\includegraphics[scale=.4]{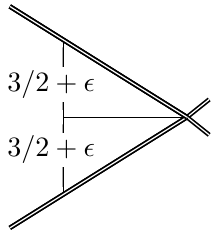}}}
\right|_{\mathrm{UV}} \!\!.
\label{eq:CYUV}
\end{equation}
Rather than evaluating the resulting UV integral directly, we observe that the same class of UV subintegrals arises from the UV limit of a topology which does not contain CY integrals, namely the topology
\begin{equation}
\mathcal{I}_{\mathrm{PL}}
=\vcenter{\hbox{\includegraphics[scale=.55]{figs/PolyLogIntegral1.pdf}}}
=\vcenter{\hbox{\includegraphics[scale=.55]{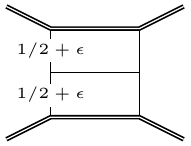}}}\,,
\label{eq:PolyLog}
\end{equation}
whose UV behavior is identical to that of \eqn{eq:CYUV}. This shows that the UV poles of $\mathcal{I}_{\mathrm{CY}}$ and $\mathcal{I}_{\mathrm{PL}}$ must be expressible in terms of the same class of special functions. However, $\mathcal{I}_{\mathrm{PL}}$ is known to evaluate entirely to multiple polylogarithms, as demonstrated, for example, in Ref.~\cite{Brammer:2025rqo}.   It follows that the ultraviolet divergent part of $\mathcal{I}_{\mathrm{CY}}$ cannot involve genuinely Calabi--Yau functions. Instead, its UV poles must be expressible solely in terms of GPLs. Thus, without performing any explicit integration, we conclude that Calabi--Yau integrals do not contribute to the ultraviolet divergent part of the 5PM amplitude.

In contrast, in the UV limit, integrals in the Heun topology depicted in \fig{fig:integrals}\subref{subfig:Heun} do not simplify in this limit. This is consistent with the fact that the Heun integrals do not cancel in the results of Refs.~\cite{Bern:2025wyd, Driesse:2026qiz}.

Note that the evaluation of the integrals considered here is simplified by the presence of triangle subloops. Nevertheless, the idea that the singular parts of integrals are simpler than the full integrals applies equally to generic four-loop integrals, even if their evaluation still requires nontrivial IBP identities and differential equations.


\Section{Conclusions}
In this Letter, we explained the absence of elliptic and Calabi--Yau integrals observed in the classical amplitude and equivalently in the conservative scattering angle at 5PM~\cite{Bern:2024adl, 
Driesse:2024xad, Driesse:2024feo, Bern:2025zno, Bern:2025wyd, Driesse:2026qiz}. A key observation is that all such contributions to the classical dynamics must be accompanied by an ultraviolet divergence. By analyzing the structure of these divergences, we showed that the more complicated special functions cannot appear in the final results.

Beyond explaining why the most intricate special functions cancel in the conservative sector of the 5PM scattering angle, our analysis suggests a more efficient alternative to the standard approach whenever the relevant physics is encoded in singular behavior.   Because ultraviolet and infrared divergences are generally easier to extract than finite terms, the method suggests a useful strategy for certain perturbative calculations in classical gravity, isolating conservative effects at even loop orders and dissipative effects at odd loop orders. 

In this work, we focused on ultraviolet singularities, which are where the most complicated functions arise at 5PM order. The complete result requires combining both ultraviolet and infrared contributions.  Although not the subject of the present paper, to confirm that such a method actually works, we collected both the infrared and ultraviolet singularities and reproduced the correct 3PM classical conservative scattering amplitude~\cite{Bern:2019nnu, Bern:2019crd}. The broader utility of the approach will be judged by its application to state-of-the-art problems, such as recent determinations of the 5PM scattering angle~\cite{Driesse:2024xad, Bern:2025wyd, Bern:2025zno, Driesse:2026qiz}.  Possible refinements include separating potential and radiation regions and extending the method to energy-loss calculations at 4PM.   Ultimately, reaching 6PM order will require new tools and ideas. Exploiting the structure of the UV and IR singularities should provide one such useful approach. 

\vskip .2 cm 

\Section{Acknowledgments}
We thank Enrico Herrmann, Radu Roiban, Volodya Smirnov, Alexander Smirnov, and Mao Zeng for numerous discussions and collaboration on related work that this paper relies on. We also thank Zeno Capatti and Bernhard Mistlberger for discussions. We thank Zeno Capatti, Enrico Herrmann, Bernhard Mistlberger, Radu Roiban and Matthias Wilhelm for comments on this manuscript.
Z.B.~is supported in part by the U.S. Department of Energy (DOE) under award number DE-SC0009937.  Z.B., A.J. and G.M. are supported in part by the European Research Council (ERC) Horizon Synergy Grant “Making Sense of the Unexpected in the Gravitational-Wave Sky” grant agreement no. GWSky-101167314.
M.R. is supported by the Department of Energy (DOE), Contract DE-AC02-76SF00515. 
The authors acknowledge the Texas Advanced Computing Center (TACC) at the University of Texas at Austin for providing high-performance computing resources that have contributed to the research results reported within this Letter. This research used resources of the National Energy Research Scientific Computing Center (NERSC), a Department of Energy User Facility using NERSC awards ERCAP 0034824 and 0037814. 
In addition, this work used computational and storage services associated with the Hoffman2 Cluster, which is operated by the UCLA Office of Advanced Research Computing’s Research Technology Group.
We are grateful to the Mani L. Bhaumik Institute for Theoretical Physics for support.


\widetext
\newpage
\section*{Supplementary material}

\subsection{Family of two-loop cuspy integrals\label{CuspyIntegrals}}
\begin{figure}[htb]
\centering
\begin{subfigure}{0.195\textwidth}
\centering
\begin{tikzpicture}[decoration={snake,segment length=2mm,amplitude=0.5mm},scale=0.75]

\coordinate (e1) at (-0.5,1.25);
\coordinate (e2) at (1.5,0.5);
\coordinate (e3) at (1.5,0.5);
\coordinate (e4) at (-0.5,-0.25);

\coordinate (v11) at (0,1);
\coordinate (v12) at (0.5,1);
\coordinate (v13) at (1,1);

\coordinate (v21) at (0,0);
\coordinate (v22) at (0.5,0);
\coordinate (v23) at (1,0);


\draw[draw=black,double,thick] (-0.5,-1.25) -- node[pos=0.3, name=n1] {} (1.5,0);
\draw[draw=black,double,thick] (-0.5,1.25)-- node[pos=0.3, name=n2] {} (1.5,0);
\draw (n2.center) arc (150:-35:0.82) node[midway, fill=white]{$2$};
\draw (n1.center) arc (35:210:-0.82) node[midway, fill=white]{$2$};

\draw[draw=black,double,thick] (-0.5,-1.25) -- (2,0.3125);
\draw[draw=black,double,thick] (-0.5,1.25) --  (2,-0.3125);

\draw[draw=black,double,thick] (-0.5,1.25) -- (1.5,0);

\end{tikzpicture}
\caption{$f_1=\Delta^2 \epsilon^2 G_{1,0,0,1,2,2,0}$}
\end{subfigure}
\begin{subfigure}{0.19\textwidth}
\begin{tikzpicture}[decoration={snake,segment length=2mm,amplitude=0.5mm},scale=0.75]

\coordinate (e1) at (-0.5,1.25);
\coordinate (e2) at (1.5,0.5);
\coordinate (e3) at (1.5,0.5);
\coordinate (e4) at (-0.5,-0.25);

\coordinate (v11) at (0,1);
\coordinate (v12) at (0.5,1);
\coordinate (v13) at (1,1);

\coordinate (v21) at (0,0);
\coordinate (v22) at (0.5,0);
\coordinate (v23) at (1,0);

\draw[draw=black,double,thick] (-0.5,-1.25) -- node[pos=0.3, name=n1] {} (1.5,0);
\draw[draw=black,double,thick] (-0.5,1.25)-- node[pos=0.3, name=n2] {} (1.5,0);
\draw (n2.center) arc (20:95:-1.35) node[midway, fill=white]{$2$};
\draw (n2.center) arc (150:-35:0.82) node[midway, fill=white]{$2$};

\draw[draw=black,double,thick] (-0.5,-1.25) -- (2,0.3125);
\draw[draw=black,double,thick] (-0.5,1.25) --  (2,-0.3125);
\draw[draw=black,double,thick] (-0.5,1.25) -- (1.5,0);

\end{tikzpicture}
\caption{$f_2=\Delta\epsilon^2 G_{1,0,0,0,0,2,2}$}
\end{subfigure}
\begin{subfigure}{0.20\textwidth}
\centering 

\begin{tikzpicture}[decoration={snake,segment length=2mm,amplitude=0.5mm},scale=0.75]

\coordinate (e1) at (-0.5,1.25);
\coordinate (e2) at (1.5,0.5);
\coordinate (e3) at (1.5,0.5);
\coordinate (e4) at (-0.5,-0.25);

\coordinate (v11) at (0,1);
\coordinate (v12) at (0.5,1);
\coordinate (v13) at (1,1);

\coordinate (v21) at (0,0);
\coordinate (v22) at (0.5,0);
\coordinate (v23) at (1,0);

\draw[draw=black,double,thick] (-0.5,-1.25) -- node[pos=0.3, name=n1] {} (1.5,0);
\draw[draw=black,double,thick] (-0.5,1.25)-- node[pos=0.3, name=n2] {} (1.5,0);
\draw (n2.center) -- (n1.center);
\draw (n2.center) arc (150:-35:0.82) node[midway, fill=white]{$2$};

\draw[draw=black,double,thick] (-0.5,-1.25) -- (2,0.3125);
\draw[draw=black,double,thick] (-0.5,1.25) --  (2,-0.3125);

\draw[draw=black,double,thick] (-0.5,1.25) -- (1.5,0);

\end{tikzpicture}

\caption{$f_3=\epsilon^3 \frac{1-x^2}{x} G_{1, 0, 0, 1, 0, 1, 2}$}
\end{subfigure}
\begin{subfigure}{0.19\textwidth}
\begin{tikzpicture}[decoration={snake,segment length=2mm,amplitude=0.5mm},scale=0.75]

\coordinate (e1) at (-0.5,1.25);
\coordinate (e2) at (1.5,0.5);
\coordinate (e3) at (1.5,0.5);
\coordinate (e4) at (-0.5,-0.25);

\coordinate (v11) at (0,1);
\coordinate (v12) at (0.5,1);
\coordinate (v13) at (1,1);

\coordinate (v21) at (0,0);
\coordinate (v22) at (0.5,0);
\coordinate (v23) at (1,0);

\draw[draw=black,double,thick] (-0.5,-1.25) -- node[pos=0.3, name=n1] {} (1.5,0);
\draw[draw=black,double,thick] (-0.5,1.25)-- node[pos=0.3, name=n2] {} (1.5,0);
\draw[draw=black] (n1.center) -- node[pos=0.5, name=n3] {} (n2.center) node[midway, fill=white]{$2$} ;
\draw (n2.center) arc (150:-35:0.82);

\draw[draw=black,double,thick] (-0.5,1.25) -- (1.5,0);

\draw[draw=black,double,thick] (-0.5,-1.25) -- (2,0.3125);
\draw[draw=black,double,thick] (-0.5,1.25) --  (2,-0.3125);
\draw[draw=white, rotate around={-30:(0.25,0.5)},fill=white] (0.5,0.4) rectangle ++(0.5,0.5);

\node at (0.8,0.5) {$2$};

\end{tikzpicture}
\caption{$f_4=\Delta \epsilon^2   G_{2,0,0,1,0,2,1}$}
\end{subfigure}
\begin{subfigure}{0.20\textwidth}
\centering
\begin{tikzpicture}[decoration={snake,segment length=2mm,amplitude=0.5mm},scale=0.75]

\coordinate (e1) at (-0.5,1.25);
\coordinate (e2) at (1.5,0.5);
\coordinate (e3) at (1.5,0.5);
\coordinate (e4) at (-0.5,-0.25);

\coordinate (v11) at (0,1);
\coordinate (v12) at (0.5,1);
\coordinate (v13) at (1,1);

\coordinate (v21) at (0,0);
\coordinate (v22) at (0.5,0);
\coordinate (v23) at (1,0);

\draw[draw=black,double,thick] (-0.5,-1.25) -- node[pos=0.3, name=n1] {} (1.5,0);
\draw[draw=black,double,thick] (-0.5,1.25)-- node[pos=0.3, name=n2] {} (1.5,0);
\draw[draw=black] (n1.center) -- node[pos=0.5, name=n3] {} (n2.center) ;
\draw[draw=black] (n3.center) -- (1.5,0);

\draw[draw=black,double,thick] (-0.5,-1.25) -- (2,0.3125);
\draw[draw=black,double,thick] (-0.5,1.25) --  (2,-0.3125);

\draw[draw=black,double,thick] (-0.5,1.25) -- (1.5,0);

\end{tikzpicture}
\caption{$f_5=\epsilon^4\frac{1-x^2}{x} G_{1,0,0,1,1,1,1}$}
\end{subfigure}
\caption{Master integrals relevant for the evaluation of Eq.~\eqref{eq:UVIntegral}.}
\label{fig:triangle_masters}
\end{figure}

Here we describe the evaluation of the integral in \eqn{eq:UVIntegral}, closely following the analysis of Ref.~\cite{Grozin:2015kna}. We write the general integral as
\begin{align}
    G_{a_1, a_2, a_3, a_4, a_5, a_6, a_7}= \int\frac{\mathrm d^D \ell_1 \mathrm d^D \ell_2}{(\imath \pi^{D/2})^2}\frac{1}{\rho_1^{a_1}\rho_2^{a_2}\rho_3^{a_3}\rho_4^{a_4}\rho_5^{a_5}\rho_6^{a_6}\rho_7^{a_7} } \,,
\end{align}
where the propagators are chosen to align with the conventions of Ref.~\cite{Parra-Martinez:2020dzs},
\begin{equation}
\begin{aligned}
\rho_1 &= 2 u_1\cdot \ell_1+\Delta\,, & \qquad 
\rho_2 & =-2 u_2\cdot \ell_1+\Delta\,, & \qquad 
\rho_3 & =-2 u_1\cdot \ell_2+\Delta\,,  & \qquad 
\rho_4 &=2 u_2\cdot \ell_2+\Delta \,,    \\ 
\rho_5 &=\ell_1^2\,, &  
\rho_6 &=\ell_2^2\,,  & 
\rho_7 & =(\ell_1+\ell_2)^2 \,.
\label{eq:props}
\end{aligned}
\end{equation}

The double triangle diagram in Eq.~\eqref{eq:UVRelation2Loop} with numerator $\mathcal N=\ell_1^2\ell_2^2$ will correspond to the integral $G_{1,0,0,1,1,1,1}$. To evaluate this, there are five relevant master integrals, depicted in Fig.~\ref{fig:triangle_masters}. 
The master integrals satisfy the differential equation,
\begin{align}
    \frac{\mathrm{d}}{\mathrm{d} x} \vec f=\epsilon\left(\frac{1}{x}A_1+\frac{1}{x+1}A_2+\frac{1}{x-1}A_3\right)\vec f \,,
\end{align}
where $(1+x^2)/2x= \sigma = u_1\cdot u_2$, and
\begin{align}
\fixTABwidth{T}
A_1=\parenMatrixstack{
0 & 0 & 0 & 0 & 0\\
0 & 0 & 0 & 0 & 0\\
0 & 0 & 0 & 1  & 0 \\
0 & -\frac12&\>\>2\>\> & 1  & 0\\
\frac12 &\frac12& 0 &-2 & -2 
}, \qquad A_2=\parenMatrixstack{
 0 & 0 & 0  & 0 & 0 \\
 0 & 0 & 0 & 0 & 0 \\
0 & 0 &-2 & 0 & 0  \\
 0 & 1 & 0 & -2   & 0  \\
0 &0 & 0 & 0  & \>\>2\>\>
},\qquad A_3=\parenMatrixstack{
 0 & 0 & 0 & 0& 0\\
 0 & 0 & 0 & 0 & 0\\
 0 & 0&\>\>2\>\> & 0 & 0 \\
0 & 0 & 0 & 0 & 0 \\
0 & 0&0 & 0  & 2
}.
\end{align}
The integrals $f_1$ and $f_2$ are independent of $x$ and can be evaluated directly \cite{Smirnov:2004ym}, and $f_3$ and $f_5$ vanish at $x=-1$ by parity. We can further fix $f_4|_{x=-1}=\frac12 f_2|_{x=-1}$ by demanding regularity at $x=-1$. Together, this provides the boundary conditions,
\begin{align}\begin{split}
    f_1|_{x=-1}&=\Delta^{-4\epsilon} \, \Gamma^2(1-\epsilon)\, \Gamma^2(1+2\epsilon)\,,\\[3pt]
    f_2|_{x=-1}&=2f_4|_{x=-1}=\Delta^{-4\epsilon} \, \Gamma^2(1-\epsilon)\, \Gamma(1+4\epsilon)\,,\\[3pt]
    f_3|_{x=-1}&=f_5|_{x=-1}=0 \,.
\end{split}
\end{align}
This system can be solved order by order in~$\epsilon$ in terms of harmonic polylogarithms~\cite{Remiddi:1999ew}. At leading order in~$\epsilon$, we compute
\begin{align}
    G_{1,0,0,1,1,1,1}(x)+G_{1,0,0,1,1,1,1}(-x)&=\frac{x\pi (\pi-\imath\ln(x))\ln(x)}{\epsilon(1-x^2)}+\mathcal O(\epsilon^0),
\end{align}
where we have chosen the branch $\ln(-x)=\ln(x)+\imath \pi$. This is equivalent to Eq.~\eqref{eq:UVIntegral} after replacing $x$ by $\sigma$.

\subsection{Integrating out subloops and numerators \label{sec:Subloops}}

\begin{figure}[htbp]
    \centering
\begin{tikzpicture}[decoration={snake,segment length=2mm,amplitude=0.5mm},scale=1.5]
\coordinate (e1) at (-0.5,1.25);
\coordinate (e2) at (1.5,1.25);
\coordinate (e3) at (1.5,-0.25);
\coordinate (e4) at (-0.5,-0.25);

\coordinate (v11) at (0,1);
\coordinate (v12) at (0.5,1);
\coordinate (v13) at (1,1);

\coordinate (v21) at (0,0.5);
\coordinate (v22) at (0.5,0.5);
\coordinate (v23) at (1,0.5);

\coordinate (v31) at (0,0);
\coordinate (v32) at (0.5,0);
\coordinate (v33) at (1,0);

\draw[draw=black] (v21) -- (v23) ;
\draw[draw=black] (v11) -- (v21) ;
\draw[draw=black] (0.65,0) -- (v23) ;
\draw[draw=black] (v13) -- (v23) ;
\draw[draw=black] (v21) -- (v31) ;
\draw[draw=black] (v21) -- (0.35,0) ;
\draw[draw=black] (v23) -- (v33) ;
\draw[draw=black,-latex] (-0.15,0.6) -> node[midway,left] {$\ell_1$}(-0.15,0.9);
\draw[draw=black,-latex] (1.15,0.1) -> node[midway,right] {$\ell_2$}(1.15,0.4) ;
\draw[draw=black,-latex] (-0.15,0.1) -> node[midway,left] {$\ell_3$}(-0.15,0.4);
\draw[draw=black,-latex] (0.2,0.6) -> node[midway,above] {$\ell_4$}(0.8,0.6) ;

\draw[draw=black,double,thick] (e1) -- (v11) -- (v13) -- (e2);
\draw[draw=black,double,thick] (e4) --(v31) -- (v33) -- (e3);

\end{tikzpicture}

\caption{Propagators and loop-momentum labels for the elliptic integral topology.}
\label{fig:elliptic}
\end{figure}
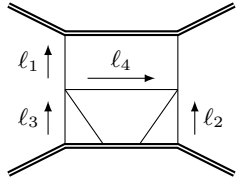

Besides being useful for determining the class of functions that can appear, the ideas explained in this Letter can be used to evaluate integrals, making use of the simplifications in infrared and ultraviolet singular regions. As a concrete example, here we explicitly evaluate the purely ultraviolet divergent sample integral of the type in Eq.~\eqref{eq:elliptic_diagram}:
\begin{align}
    \mathcal I_{\rm ell}=\int\frac{\mathrm{d}^D\ell_1 \, \mathrm{d}^D\ell_2 \, \mathrm{d}^D\ell_3\, \mathrm{d}^D\ell_4}{(\imath\pi^{D/2})^{4}} \, \frac{\deltabar(2u_2\cdot\ell_2)\, \deltabar(-2u_2\cdot\ell_3)\, \deltabar(2u_1\cdot\ell_1)\, \deltabar(-2u_2\cdot(\ell_4+\ell_1))\, \mathcal{N}}{\ell_1^2\ell_2^2\ell_3^2\ell_4^2 (\ell_1-q)^2(\ell_1+\ell_2+\ell_4-q)^2(\ell_3-\ell_1-\ell_4)^2}\, ,\label{eq:elliptic}
\end{align}
where the loop momenta correspond to the labels in \fig{fig:elliptic} and $q$ is the momentum transfer between the two massive particles. For this example, we choose a numerator that appears in a four-loop  classical gravity calculation~\cite{Bern:2025wyd},
\begin{align}
    \mathcal{N} = -\ell_4\cdot(\ell_1+\ell_2+\ell_4-q)\,.
\end{align}
We first directly evaluate the integrals over $\ell_3$ and $\ell_2$, which, diagrammatically, correspond to integrating out the two triangle subloops. For the numerator $\mathcal{N}$ independent of $\ell_3$, the integral over $\ell_3$ is computed in Ref.~\cite{Smirnov:2004ym} with the result:
\begin{align}\label{left_triangle_1}
    \mathcal I_{\rm left} =\int\frac{\mathrm{d}^D\ell_3}{\imath\pi^{D/2}}\, \frac{\deltabar(-2u_2\cdot \ell_3)}{\ell_3^2(\ell_3-\ell_1-\ell_4)^2} 
    =\frac{\Gamma^2(\frac{1}{2}-\epsilon) \Gamma(\frac{1}{2}+\epsilon)\Gamma(\frac{1}{2})}{{\Gamma(1-2\epsilon)}} \, \frac{1}{(-(\ell_1+\ell_4)^2)^{\frac{1}{2}+\epsilon}}\,.
\end{align}
The integral over $\ell_2$ is given by
\begin{align}\label{right_triangle}
    \mathcal I_{\rm right}= \deltabar(-2u_2\cdot (\ell_1+\ell_4))\int\frac{\mathrm{d}^D\ell_2 }{\imath\pi^{D/2}} \, \frac{ \ell_4\cdot(\ell_2+\ell_1+\ell_4-q)\, \deltabar(2u_2\cdot \ell_2) }{\ell_2^2(\ell_2+\ell_1+\ell_4-q)^2}\,.
\end{align}
The term containing $\ell_4\cdot\ell_2$ in the numerator can be simplified using Passarino-Veltman reduction. Keeping only the terms that are not canceled by the cut matter line, $\deltabar(-2u_2\cdot (\ell_1+\ell_4))$, we find the resulting integral to be
\begin{align}\label{right_triangle_2}
     \mathcal I_{\rm right}=\frac{1}{2}\ell_4\cdot(\ell_1+\ell_4-q) \, \deltabar(-2u_2\cdot (\ell_1+\ell_4))\int\frac{\mathrm{d}^D\ell_2}{\imath\pi^{D/2}}\, \frac{\deltabar(2u_2\cdot \ell_2)}{\ell_2^2(\ell_2+\ell_1+\ell_4-q)^2} \,.
    \end{align}
This is an integral of the same type as Eq.~\eqref{left_triangle_1}, which evaluates to,
\begin{align}
     \mathcal I_{\rm right}=\frac{1}{2}\ell_4\cdot(\ell_1+\ell_4-q) \, \deltabar(-2u_2\cdot (\ell_1+\ell_4))\, \frac{\Gamma^2(\frac{1}{2}-\epsilon)\Gamma(\frac{1}{2}+\epsilon)\Gamma(\frac{1}{2})}{{\Gamma(1-2\epsilon)}}\, \frac{1}{(-(\ell_1+\ell_4-q)^2)^{\frac{1}{2}+\epsilon}} \,.
\end{align}
Using the results of the subloop integration and performing the change of variables, $\ell_1\to \ell_1,\ell_4 \to q-\ell_1-\ell_2$, we can rewrite the integral as
\begin{align}
    \mathcal I_{\rm ell}=-\frac{1}{2}\biggl(\frac{\Gamma^2(\frac{1}{2}-\epsilon) \Gamma(\frac{1}{2}+\epsilon)\Gamma(\frac{1}{2})}{{\Gamma(1-2\epsilon)}}\biggr)^2 \int\frac{\mathrm{d}^D\ell_1 \, \mathrm{d}^D\ell_2}{(\imath\pi^{D/2})^{2}}\, \frac{\ell_2\cdot(\ell_1+\ell_2-q)\deltabar(2u_1\cdot\ell_1)\, \deltabar(2u_2\cdot\ell_2)}{\ell_1^2(-\ell_2^2)^{\frac{1}{2}+\epsilon}(\ell_1-q)^2(-(\ell_2-q)^2)^{\frac{1}{2}+\epsilon}(\ell_1+\ell_2-q)^2}\,.
\end{align}
The UV limit of this integral to leading order in $\epsilon$ is
\begin{align}
\left.\mathcal{I}_{\mathrm{ell}}\right|_{\mathrm{UV}}={}&\left.\frac{1}{2}\pi^4\int\frac{\mathrm{d}^D\ell_1\mathrm{d}^D\ell_2} {(\imath\pi^{D/2})^2}\, \frac{\ell_2\cdot(\ell_1+\ell_2)\deltabar(2u_1\cdot \ell_1{+}\Delta)\, \deltabar(2u_2\cdot \ell_2{+}\Delta)\,}{[\ell_1^2]^{2}\ell_2^2(\ell_1{+}\ell_2)^2}\right|_{\mathrm{UV}}.\nonumber
\end{align}
This expression is equivalent to Eq.~\eqref{eq:UVH1} with $\mathcal{N}=\frac{1}{2}\pi^4\,\ell_2\cdot(\ell_1+\ell_2)\,\ell_2^2$. Evaluating this integral in the UV limit gives,
\begin{align}
    \left.\mathcal{I}_{\mathrm{ell}}\right|_{\mathrm{UV}} = \frac{\pi^6 {\operatorname{ arccosh}}(\sigma)}{8\,\epsilon\,\sqrt{\sigma^2-1}}-\frac{\imath \pi^5(\sigma^2 -1+  {\operatorname{ arccosh}}^2(\sigma))}{8 \epsilon \sqrt{\sigma^2-1}}+\mathcal{O}(\epsilon^0)\,.
\end{align}
The real part of the integral is relevant for calculating conservative dynamics.  We have verified that this agrees with the value obtained via integration by parts, following the setup in Ref.~\cite{Bern:2024adl} for the conservative part.

\bibliographystyle{apsrev4-1}
\bibliography{CY_cancel}

\begin{thebibliography}{110}%
\makeatletter
\providecommand \@ifxundefined [1]{%
 \@ifx{#1\undefined}
}%
\providecommand \@ifnum [1]{%
 \ifnum #1\expandafter \@firstoftwo
 \else \expandafter \@secondoftwo
 \fi
}%
\providecommand \@ifx [1]{%
 \ifx #1\expandafter \@firstoftwo
 \else \expandafter \@secondoftwo
 \fi
}%
\providecommand \natexlab [1]{#1}%
\providecommand \enquote  [1]{``#1''}%
\providecommand \bibnamefont  [1]{#1}%
\providecommand \bibfnamefont [1]{#1}%
\providecommand \citenamefont [1]{#1}%
\providecommand \href@noop [0]{\@secondoftwo}%
\providecommand \href [0]{\begingroup \@sanitize@url \@href}%
\providecommand \@href[1]{\@@startlink{#1}\@@href}%
\providecommand \@@href[1]{\endgroup#1\@@endlink}%
\providecommand \@sanitize@url [0]{\catcode `\\12\catcode `\$12\catcode `\&12\catcode `\#12\catcode `\^12\catcode `\_12\catcode `\%12\relax}%
\providecommand \@@startlink[1]{}%
\providecommand \@@endlink[0]{}%
\providecommand \url  [0]{\begingroup\@sanitize@url \@url }%
\providecommand \@url [1]{\endgroup\@href {#1}{\urlprefix }}%
\providecommand \urlprefix  [0]{URL }%
\providecommand \Eprint [0]{\href }%
\providecommand \doibase [0]{http://dx.doi.org/}%
\providecommand \selectlanguage [0]{\@gobble}%
\providecommand \bibinfo  [0]{\@secondoftwo}%
\providecommand \bibfield  [0]{\@secondoftwo}%
\providecommand \translation [1]{[#1]}%
\providecommand \BibitemOpen [0]{}%
\providecommand \bibitemStop [0]{}%
\providecommand \bibitemNoStop [0]{.\EOS\space}%
\providecommand \EOS [0]{\spacefactor3000\relax}%
\providecommand \BibitemShut  [1]{\csname bibitem#1\endcsname}%
\let\auto@bib@innerbib\@empty
\bibitem [{\citenamefont {Punturo}\ \emph {et~al.}(2010)\citenamefont {Punturo} \emph {et~al.}}]{Punturo:2010zz}%
  \BibitemOpen
  \bibfield  {author} {\bibinfo {author} {\bibfnamefont {M.}~\bibnamefont {Punturo}} \emph {et~al.},\ }\href {\doibase 10.1088/0264-9381/27/19/194002} {\bibfield  {journal} {\bibinfo  {journal} {Class. Quant. Grav.}\ }\textbf {\bibinfo {volume} {27}},\ \bibinfo {pages} {194002} (\bibinfo {year} {2010})}\BibitemShut {NoStop}%
\bibitem [{\citenamefont {Colpi}\ \emph {et~al.}(2024)\citenamefont {Colpi} \emph {et~al.}}]{LISA:2024hlh}%
  \BibitemOpen
  \bibfield  {author} {\bibinfo {author} {\bibfnamefont {M.}~\bibnamefont {Colpi}} \emph {et~al.} (\bibinfo {collaboration} {LISA}),\ }\href@noop {} {\  (\bibinfo {year} {2024})},\ \Eprint {http://arxiv.org/abs/2402.07571} {arXiv:2402.07571 [astro-ph.CO]} \BibitemShut {NoStop}%
\bibitem [{\citenamefont {Reitze}\ \emph {et~al.}(2019)\citenamefont {Reitze} \emph {et~al.}}]{Reitze:2019iox}%
  \BibitemOpen
  \bibfield  {author} {\bibinfo {author} {\bibfnamefont {D.}~\bibnamefont {Reitze}} \emph {et~al.},\ }\href@noop {} {\bibfield  {journal} {\bibinfo  {journal} {Bull. Am. Astron. Soc.}\ }\textbf {\bibinfo {volume} {51}},\ \bibinfo {pages} {035} (\bibinfo {year} {2019})},\ \Eprint {http://arxiv.org/abs/1907.04833} {arXiv:1907.04833 [astro-ph.IM]} \BibitemShut {NoStop}%
\bibitem [{\citenamefont {Fritschel}\ \emph {et~al.}()\citenamefont {Fritschel} \emph {et~al.}}]{LIGOasharp}%
  \BibitemOpen
  \bibfield  {author} {\bibinfo {author} {\bibfnamefont {P.}~\bibnamefont {Fritschel}} \emph {et~al.},\ }\href@noop {} {\enquote {\bibinfo {title} {{Report from the LSC Post-O5 Study Group}},}\ }\bibinfo {note} {Tech. Rep. T2200287 (LIGO, 2022)}\BibitemShut {NoStop}%
\bibitem [{\citenamefont {Abac}\ \emph {et~al.}(2025)\citenamefont {Abac} \emph {et~al.}}]{Abac:2025saz}%
  \BibitemOpen
  \bibfield  {author} {\bibinfo {author} {\bibfnamefont {A.}~\bibnamefont {Abac}} \emph {et~al.},\ }\href@noop {} {\  (\bibinfo {year} {2025})},\ \Eprint {http://arxiv.org/abs/2503.12263} {arXiv:2503.12263 [gr-qc]} \BibitemShut {NoStop}%
\bibitem [{\citenamefont {Pretorius}(2005)}]{Pretorius:2005gq}%
  \BibitemOpen
  \bibfield  {author} {\bibinfo {author} {\bibfnamefont {F.}~\bibnamefont {Pretorius}},\ }\href {\doibase 10.1103/PhysRevLett.95.121101} {\bibfield  {journal} {\bibinfo  {journal} {Phys. Rev. Lett.}\ }\textbf {\bibinfo {volume} {95}},\ \bibinfo {pages} {121101} (\bibinfo {year} {2005})},\ \Eprint {http://arxiv.org/abs/gr-qc/0507014} {arXiv:gr-qc/0507014} \BibitemShut {NoStop}%
\bibitem [{\citenamefont {Campanelli}\ \emph {et~al.}(2006)\citenamefont {Campanelli}, \citenamefont {Lousto}, \citenamefont {Marronetti},\ and\ \citenamefont {Zlochower}}]{Campanelli:2005dd}%
  \BibitemOpen
  \bibfield  {author} {\bibinfo {author} {\bibfnamefont {M.}~\bibnamefont {Campanelli}}, \bibinfo {author} {\bibfnamefont {C.~O.}\ \bibnamefont {Lousto}}, \bibinfo {author} {\bibfnamefont {P.}~\bibnamefont {Marronetti}}, \ and\ \bibinfo {author} {\bibfnamefont {Y.}~\bibnamefont {Zlochower}},\ }\href {\doibase 10.1103/PhysRevLett.96.111101} {\bibfield  {journal} {\bibinfo  {journal} {Phys. Rev. Lett.}\ }\textbf {\bibinfo {volume} {96}},\ \bibinfo {pages} {111101} (\bibinfo {year} {2006})},\ \Eprint {http://arxiv.org/abs/gr-qc/0511048} {arXiv:gr-qc/0511048} \BibitemShut {NoStop}%
\bibitem [{\citenamefont {Baker}\ \emph {et~al.}(2006)\citenamefont {Baker}, \citenamefont {Centrella}, \citenamefont {Choi}, \citenamefont {Koppitz},\ and\ \citenamefont {van Meter}}]{Baker:2005vv}%
  \BibitemOpen
  \bibfield  {author} {\bibinfo {author} {\bibfnamefont {J.~G.}\ \bibnamefont {Baker}}, \bibinfo {author} {\bibfnamefont {J.}~\bibnamefont {Centrella}}, \bibinfo {author} {\bibfnamefont {D.-I.}\ \bibnamefont {Choi}}, \bibinfo {author} {\bibfnamefont {M.}~\bibnamefont {Koppitz}}, \ and\ \bibinfo {author} {\bibfnamefont {J.}~\bibnamefont {van Meter}},\ }\href {\doibase 10.1103/PhysRevLett.96.111102} {\bibfield  {journal} {\bibinfo  {journal} {Phys. Rev. Lett.}\ }\textbf {\bibinfo {volume} {96}},\ \bibinfo {pages} {111102} (\bibinfo {year} {2006})},\ \Eprint {http://arxiv.org/abs/gr-qc/0511103} {arXiv:gr-qc/0511103} \BibitemShut {NoStop}%
\bibitem [{\citenamefont {Damour}\ \emph {et~al.}(2014)\citenamefont {Damour}, \citenamefont {Guercilena}, \citenamefont {Hinder}, \citenamefont {Hopper}, \citenamefont {Nagar},\ and\ \citenamefont {Rezzolla}}]{Damour:2014afa}%
  \BibitemOpen
  \bibfield  {author} {\bibinfo {author} {\bibfnamefont {T.}~\bibnamefont {Damour}}, \bibinfo {author} {\bibfnamefont {F.}~\bibnamefont {Guercilena}}, \bibinfo {author} {\bibfnamefont {I.}~\bibnamefont {Hinder}}, \bibinfo {author} {\bibfnamefont {S.}~\bibnamefont {Hopper}}, \bibinfo {author} {\bibfnamefont {A.}~\bibnamefont {Nagar}}, \ and\ \bibinfo {author} {\bibfnamefont {L.}~\bibnamefont {Rezzolla}},\ }\href {\doibase 10.1103/PhysRevD.89.081503} {\bibfield  {journal} {\bibinfo  {journal} {Phys. Rev. D}\ }\textbf {\bibinfo {volume} {89}},\ \bibinfo {pages} {081503} (\bibinfo {year} {2014})},\ \Eprint {http://arxiv.org/abs/1402.7307} {arXiv:1402.7307 [gr-qc]} \BibitemShut {NoStop}%
\bibitem [{\citenamefont {Mino}\ \emph {et~al.}(1997)\citenamefont {Mino}, \citenamefont {Sasaki},\ and\ \citenamefont {Tanaka}}]{Mino:1996nk}%
  \BibitemOpen
  \bibfield  {author} {\bibinfo {author} {\bibfnamefont {Y.}~\bibnamefont {Mino}}, \bibinfo {author} {\bibfnamefont {M.}~\bibnamefont {Sasaki}}, \ and\ \bibinfo {author} {\bibfnamefont {T.}~\bibnamefont {Tanaka}},\ }\href {\doibase 10.1103/PhysRevD.55.3457} {\bibfield  {journal} {\bibinfo  {journal} {Phys. Rev. D}\ }\textbf {\bibinfo {volume} {55}},\ \bibinfo {pages} {3457} (\bibinfo {year} {1997})},\ \Eprint {http://arxiv.org/abs/gr-qc/9606018} {arXiv:gr-qc/9606018} \BibitemShut {NoStop}%
\bibitem [{\citenamefont {Quinn}\ and\ \citenamefont {Wald}(1997)}]{Quinn:1996am}%
  \BibitemOpen
  \bibfield  {author} {\bibinfo {author} {\bibfnamefont {T.~C.}\ \bibnamefont {Quinn}}\ and\ \bibinfo {author} {\bibfnamefont {R.~M.}\ \bibnamefont {Wald}},\ }\href {\doibase 10.1103/PhysRevD.56.3381} {\bibfield  {journal} {\bibinfo  {journal} {Phys. Rev. D}\ }\textbf {\bibinfo {volume} {56}},\ \bibinfo {pages} {3381} (\bibinfo {year} {1997})},\ \Eprint {http://arxiv.org/abs/gr-qc/9610053} {arXiv:gr-qc/9610053} \BibitemShut {NoStop}%
\bibitem [{\citenamefont {Poisson}\ \emph {et~al.}(2011)\citenamefont {Poisson}, \citenamefont {Pound},\ and\ \citenamefont {Vega}}]{Poisson:2011nh}%
  \BibitemOpen
  \bibfield  {author} {\bibinfo {author} {\bibfnamefont {E.}~\bibnamefont {Poisson}}, \bibinfo {author} {\bibfnamefont {A.}~\bibnamefont {Pound}}, \ and\ \bibinfo {author} {\bibfnamefont {I.}~\bibnamefont {Vega}},\ }\href {\doibase 10.12942/lrr-2011-7} {\bibfield  {journal} {\bibinfo  {journal} {Living Rev. Rel.}\ }\textbf {\bibinfo {volume} {14}},\ \bibinfo {pages} {7} (\bibinfo {year} {2011})},\ \Eprint {http://arxiv.org/abs/1102.0529} {arXiv:1102.0529 [gr-qc]} \BibitemShut {NoStop}%
\bibitem [{\citenamefont {Barack}\ and\ \citenamefont {Pound}(2019)}]{Barack:2018yvs}%
  \BibitemOpen
  \bibfield  {author} {\bibinfo {author} {\bibfnamefont {L.}~\bibnamefont {Barack}}\ and\ \bibinfo {author} {\bibfnamefont {A.}~\bibnamefont {Pound}},\ }\href {\doibase 10.1088/1361-6633/aae552} {\bibfield  {journal} {\bibinfo  {journal} {Rept. Prog. Phys.}\ }\textbf {\bibinfo {volume} {82}},\ \bibinfo {pages} {016904} (\bibinfo {year} {2019})},\ \Eprint {http://arxiv.org/abs/1805.10385} {arXiv:1805.10385 [gr-qc]} \BibitemShut {NoStop}%
\bibitem [{\citenamefont {Goldberger}\ and\ \citenamefont {Rothstein}(2006)}]{Goldberger:2004jt}%
  \BibitemOpen
  \bibfield  {author} {\bibinfo {author} {\bibfnamefont {W.~D.}\ \bibnamefont {Goldberger}}\ and\ \bibinfo {author} {\bibfnamefont {I.~Z.}\ \bibnamefont {Rothstein}},\ }\href {\doibase 10.1103/PhysRevD.73.104029} {\bibfield  {journal} {\bibinfo  {journal} {Phys. Rev. D}\ }\textbf {\bibinfo {volume} {73}},\ \bibinfo {pages} {104029} (\bibinfo {year} {2006})},\ \Eprint {http://arxiv.org/abs/hep-th/0409156} {arXiv:hep-th/0409156} \BibitemShut {NoStop}%
\bibitem [{\citenamefont {Cheung}\ \emph {et~al.}(2018)\citenamefont {Cheung}, \citenamefont {Rothstein},\ and\ \citenamefont {Solon}}]{Cheung:2018wkq}%
  \BibitemOpen
  \bibfield  {author} {\bibinfo {author} {\bibfnamefont {C.}~\bibnamefont {Cheung}}, \bibinfo {author} {\bibfnamefont {I.~Z.}\ \bibnamefont {Rothstein}}, \ and\ \bibinfo {author} {\bibfnamefont {M.~P.}\ \bibnamefont {Solon}},\ }\href {\doibase 10.1103/PhysRevLett.121.251101} {\bibfield  {journal} {\bibinfo  {journal} {Phys. Rev. Lett.}\ }\textbf {\bibinfo {volume} {121}},\ \bibinfo {pages} {251101} (\bibinfo {year} {2018})},\ \Eprint {http://arxiv.org/abs/1808.02489} {arXiv:1808.02489 [hep-th]} \BibitemShut {NoStop}%
\bibitem [{\citenamefont {Droste}(1916)}]{Droste:1916}%
  \BibitemOpen
  \bibfield  {author} {\bibinfo {author} {\bibfnamefont {J.}~\bibnamefont {Droste}},\ }\href@noop {} {\bibfield  {journal} {\bibinfo  {journal} {Proc. Acad. Sci. Amst.}\ }\textbf {\bibinfo {volume} {19}},\ \bibinfo {pages} {447} (\bibinfo {year} {1916})}\BibitemShut {NoStop}%
\bibitem [{\citenamefont {Lorentz}\ and\ \citenamefont {Droste}(1917)}]{Droste:1917}%
  \BibitemOpen
  \bibfield  {author} {\bibinfo {author} {\bibfnamefont {H.}~\bibnamefont {Lorentz}}\ and\ \bibinfo {author} {\bibfnamefont {J.}~\bibnamefont {Droste}},\ }\href@noop {} {\bibfield  {journal} {\bibinfo  {journal} {Verslagen der Afdeeling Natuurkunde van de Koninklijke Akademie van Wetenschappen}\ }\textbf {\bibinfo {volume} {26}},\ \bibinfo {pages} {392} (\bibinfo {year} {1917})}\BibitemShut {NoStop}%
\bibitem [{\citenamefont {Einstein}\ \emph {et~al.}(1938)\citenamefont {Einstein}, \citenamefont {Infeld},\ and\ \citenamefont {Hoffmann}}]{Einstein:1938yz}%
  \BibitemOpen
  \bibfield  {author} {\bibinfo {author} {\bibfnamefont {A.}~\bibnamefont {Einstein}}, \bibinfo {author} {\bibfnamefont {L.}~\bibnamefont {Infeld}}, \ and\ \bibinfo {author} {\bibfnamefont {B.}~\bibnamefont {Hoffmann}},\ }\href {\doibase 10.2307/1968714} {\bibfield  {journal} {\bibinfo  {journal} {Annals Math.}\ }\textbf {\bibinfo {volume} {39}},\ \bibinfo {pages} {65} (\bibinfo {year} {1938})}\BibitemShut {NoStop}%
\bibitem [{\citenamefont {Ohta}\ \emph {et~al.}(1973)\citenamefont {Ohta}, \citenamefont {Okamura}, \citenamefont {Kimura},\ and\ \citenamefont {Hiida}}]{Ohta:1973je}%
  \BibitemOpen
  \bibfield  {author} {\bibinfo {author} {\bibfnamefont {T.}~\bibnamefont {Ohta}}, \bibinfo {author} {\bibfnamefont {H.}~\bibnamefont {Okamura}}, \bibinfo {author} {\bibfnamefont {T.}~\bibnamefont {Kimura}}, \ and\ \bibinfo {author} {\bibfnamefont {K.}~\bibnamefont {Hiida}},\ }\href {\doibase 10.1143/PTP.50.492} {\bibfield  {journal} {\bibinfo  {journal} {Prog. Theor. Phys.}\ }\textbf {\bibinfo {volume} {50}},\ \bibinfo {pages} {492} (\bibinfo {year} {1973})}\BibitemShut {NoStop}%
\bibitem [{\citenamefont {Blanchet}(2014)}]{Blanchet:2013haa}%
  \BibitemOpen
  \bibfield  {author} {\bibinfo {author} {\bibfnamefont {L.}~\bibnamefont {Blanchet}},\ }\href {\doibase 10.12942/lrr-2014-2} {\bibfield  {journal} {\bibinfo  {journal} {Living Rev. Rel.}\ }\textbf {\bibinfo {volume} {17}},\ \bibinfo {pages} {2} (\bibinfo {year} {2014})},\ \Eprint {http://arxiv.org/abs/1310.1528} {arXiv:1310.1528 [gr-qc]} \BibitemShut {NoStop}%
\bibitem [{\citenamefont {Bertotti}(1956)}]{Bertotti:1956pxu}%
  \BibitemOpen
  \bibfield  {author} {\bibinfo {author} {\bibfnamefont {B.}~\bibnamefont {Bertotti}},\ }\href {\doibase 10.1007/bf02746175} {\bibfield  {journal} {\bibinfo  {journal} {Nuovo Cim.}\ }\textbf {\bibinfo {volume} {4}},\ \bibinfo {pages} {898} (\bibinfo {year} {1956})}\BibitemShut {NoStop}%
\bibitem [{\citenamefont {Kerr}(1959)}]{Kerr:1959zlt}%
  \BibitemOpen
  \bibfield  {author} {\bibinfo {author} {\bibfnamefont {R.~P.}\ \bibnamefont {Kerr}},\ }\href {\doibase 10.1007/bf02732767} {\bibfield  {journal} {\bibinfo  {journal} {Nuovo Cim.}\ }\textbf {\bibinfo {volume} {13}},\ \bibinfo {pages} {469} (\bibinfo {year} {1959})}\BibitemShut {NoStop}%
\bibitem [{\citenamefont {Bertotti}\ and\ \citenamefont {Plebanski}(1960)}]{Bertotti:1960wuq}%
  \BibitemOpen
  \bibfield  {author} {\bibinfo {author} {\bibfnamefont {B.}~\bibnamefont {Bertotti}}\ and\ \bibinfo {author} {\bibfnamefont {J.}~\bibnamefont {Plebanski}},\ }\href {\doibase 10.1016/0003-4916(60)90132-9} {\bibfield  {journal} {\bibinfo  {journal} {Annals Phys.}\ }\textbf {\bibinfo {volume} {11}},\ \bibinfo {pages} {169} (\bibinfo {year} {1960})}\BibitemShut {NoStop}%
\bibitem [{\citenamefont {Westpfahl}\ and\ \citenamefont {Goller}(1979)}]{Westpfahl:1979gu}%
  \BibitemOpen
  \bibfield  {author} {\bibinfo {author} {\bibfnamefont {K.}~\bibnamefont {Westpfahl}}\ and\ \bibinfo {author} {\bibfnamefont {M.}~\bibnamefont {Goller}},\ }\href {\doibase 10.1007/BF02817047} {\bibfield  {journal} {\bibinfo  {journal} {Lett. Nuovo Cim.}\ }\textbf {\bibinfo {volume} {26}},\ \bibinfo {pages} {573} (\bibinfo {year} {1979})}\BibitemShut {NoStop}%
\bibitem [{\citenamefont {Portilla}(1980)}]{Portilla:1980uz}%
  \BibitemOpen
  \bibfield  {author} {\bibinfo {author} {\bibfnamefont {M.}~\bibnamefont {Portilla}},\ }\href {\doibase 10.1088/0305-4470/13/12/017} {\bibfield  {journal} {\bibinfo  {journal} {J. Phys. A}\ }\textbf {\bibinfo {volume} {13}},\ \bibinfo {pages} {3677} (\bibinfo {year} {1980})}\BibitemShut {NoStop}%
\bibitem [{\citenamefont {Bel}\ \emph {et~al.}(1981)\citenamefont {Bel}, \citenamefont {Damour}, \citenamefont {Deruelle}, \citenamefont {Ibanez},\ and\ \citenamefont {Martin}}]{Bel:1981be}%
  \BibitemOpen
  \bibfield  {author} {\bibinfo {author} {\bibfnamefont {L.}~\bibnamefont {Bel}}, \bibinfo {author} {\bibfnamefont {T.}~\bibnamefont {Damour}}, \bibinfo {author} {\bibfnamefont {N.}~\bibnamefont {Deruelle}}, \bibinfo {author} {\bibfnamefont {J.}~\bibnamefont {Ibanez}}, \ and\ \bibinfo {author} {\bibfnamefont {J.}~\bibnamefont {Martin}},\ }\href {\doibase 10.1007/BF00756073} {\bibfield  {journal} {\bibinfo  {journal} {Gen. Rel. Grav.}\ }\textbf {\bibinfo {volume} {13}},\ \bibinfo {pages} {963} (\bibinfo {year} {1981})}\BibitemShut {NoStop}%
\bibitem [{\citenamefont {Bern}\ \emph {et~al.}(2019{\natexlab{a}})\citenamefont {Bern}, \citenamefont {Cheung}, \citenamefont {Roiban}, \citenamefont {Shen}, \citenamefont {Solon},\ and\ \citenamefont {Zeng}}]{Bern:2019nnu}%
  \BibitemOpen
  \bibfield  {author} {\bibinfo {author} {\bibfnamefont {Z.}~\bibnamefont {Bern}}, \bibinfo {author} {\bibfnamefont {C.}~\bibnamefont {Cheung}}, \bibinfo {author} {\bibfnamefont {R.}~\bibnamefont {Roiban}}, \bibinfo {author} {\bibfnamefont {C.-H.}\ \bibnamefont {Shen}}, \bibinfo {author} {\bibfnamefont {M.~P.}\ \bibnamefont {Solon}}, \ and\ \bibinfo {author} {\bibfnamefont {M.}~\bibnamefont {Zeng}},\ }\href {\doibase 10.1103/PhysRevLett.122.201603} {\bibfield  {journal} {\bibinfo  {journal} {Phys. Rev. Lett.}\ }\textbf {\bibinfo {volume} {122}},\ \bibinfo {pages} {201603} (\bibinfo {year} {2019}{\natexlab{a}})},\ \Eprint {http://arxiv.org/abs/1901.04424} {arXiv:1901.04424 [hep-th]} \BibitemShut {NoStop}%
\bibitem [{\citenamefont {Bern}\ \emph {et~al.}(2019{\natexlab{b}})\citenamefont {Bern}, \citenamefont {Cheung}, \citenamefont {Roiban}, \citenamefont {Shen}, \citenamefont {Solon},\ and\ \citenamefont {Zeng}}]{Bern:2019crd}%
  \BibitemOpen
  \bibfield  {author} {\bibinfo {author} {\bibfnamefont {Z.}~\bibnamefont {Bern}}, \bibinfo {author} {\bibfnamefont {C.}~\bibnamefont {Cheung}}, \bibinfo {author} {\bibfnamefont {R.}~\bibnamefont {Roiban}}, \bibinfo {author} {\bibfnamefont {C.-H.}\ \bibnamefont {Shen}}, \bibinfo {author} {\bibfnamefont {M.~P.}\ \bibnamefont {Solon}}, \ and\ \bibinfo {author} {\bibfnamefont {M.}~\bibnamefont {Zeng}},\ }\href {\doibase 10.1007/JHEP10(2019)206} {\bibfield  {journal} {\bibinfo  {journal} {JHEP}\ }\textbf {\bibinfo {volume} {10}},\ \bibinfo {pages} {206} (\bibinfo {year} {2019}{\natexlab{b}})},\ \Eprint {http://arxiv.org/abs/1908.01493} {arXiv:1908.01493 [hep-th]} \BibitemShut {NoStop}%
\bibitem [{\citenamefont {Bern}\ \emph {et~al.}(2021)\citenamefont {Bern}, \citenamefont {Parra-Martinez}, \citenamefont {Roiban}, \citenamefont {Ruf}, \citenamefont {Shen}, \citenamefont {Solon},\ and\ \citenamefont {Zeng}}]{Bern:2021dqo}%
  \BibitemOpen
  \bibfield  {author} {\bibinfo {author} {\bibfnamefont {Z.}~\bibnamefont {Bern}}, \bibinfo {author} {\bibfnamefont {J.}~\bibnamefont {Parra-Martinez}}, \bibinfo {author} {\bibfnamefont {R.}~\bibnamefont {Roiban}}, \bibinfo {author} {\bibfnamefont {M.~S.}\ \bibnamefont {Ruf}}, \bibinfo {author} {\bibfnamefont {C.-H.}\ \bibnamefont {Shen}}, \bibinfo {author} {\bibfnamefont {M.~P.}\ \bibnamefont {Solon}}, \ and\ \bibinfo {author} {\bibfnamefont {M.}~\bibnamefont {Zeng}},\ }\href {\doibase 10.1103/PhysRevLett.126.171601} {\bibfield  {journal} {\bibinfo  {journal} {Phys. Rev. Lett.}\ }\textbf {\bibinfo {volume} {126}},\ \bibinfo {pages} {171601} (\bibinfo {year} {2021})},\ \Eprint {http://arxiv.org/abs/2101.07254} {arXiv:2101.07254 [hep-th]} \BibitemShut {NoStop}%
\bibitem [{\citenamefont {Bern}\ \emph {et~al.}(2022)\citenamefont {Bern}, \citenamefont {Parra-Martinez}, \citenamefont {Roiban}, \citenamefont {Ruf}, \citenamefont {Shen}, \citenamefont {Solon},\ and\ \citenamefont {Zeng}}]{Bern:2021yeh}%
  \BibitemOpen
  \bibfield  {author} {\bibinfo {author} {\bibfnamefont {Z.}~\bibnamefont {Bern}}, \bibinfo {author} {\bibfnamefont {J.}~\bibnamefont {Parra-Martinez}}, \bibinfo {author} {\bibfnamefont {R.}~\bibnamefont {Roiban}}, \bibinfo {author} {\bibfnamefont {M.~S.}\ \bibnamefont {Ruf}}, \bibinfo {author} {\bibfnamefont {C.-H.}\ \bibnamefont {Shen}}, \bibinfo {author} {\bibfnamefont {M.~P.}\ \bibnamefont {Solon}}, \ and\ \bibinfo {author} {\bibfnamefont {M.}~\bibnamefont {Zeng}},\ }\href {\doibase 10.1103/PhysRevLett.128.161103} {\bibfield  {journal} {\bibinfo  {journal} {Phys. Rev. Lett.}\ }\textbf {\bibinfo {volume} {128}},\ \bibinfo {pages} {161103} (\bibinfo {year} {2022})},\ \Eprint {http://arxiv.org/abs/2112.10750} {arXiv:2112.10750 [hep-th]} \BibitemShut {NoStop}%
\bibitem [{\citenamefont {Dlapa}\ \emph {et~al.}(2022{\natexlab{a}})\citenamefont {Dlapa}, \citenamefont {K\"alin}, \citenamefont {Liu},\ and\ \citenamefont {Porto}}]{Dlapa:2021npj}%
  \BibitemOpen
  \bibfield  {author} {\bibinfo {author} {\bibfnamefont {C.}~\bibnamefont {Dlapa}}, \bibinfo {author} {\bibfnamefont {G.}~\bibnamefont {K\"alin}}, \bibinfo {author} {\bibfnamefont {Z.}~\bibnamefont {Liu}}, \ and\ \bibinfo {author} {\bibfnamefont {R.~A.}\ \bibnamefont {Porto}},\ }\href {\doibase 10.1016/j.physletb.2022.137203} {\bibfield  {journal} {\bibinfo  {journal} {Phys. Lett. B}\ }\textbf {\bibinfo {volume} {831}},\ \bibinfo {pages} {137203} (\bibinfo {year} {2022}{\natexlab{a}})},\ \Eprint {http://arxiv.org/abs/2106.08276} {arXiv:2106.08276 [hep-th]} \BibitemShut {NoStop}%
\bibitem [{\citenamefont {Dlapa}\ \emph {et~al.}(2022{\natexlab{b}})\citenamefont {Dlapa}, \citenamefont {K\"alin}, \citenamefont {Liu},\ and\ \citenamefont {Porto}}]{Dlapa:2021vgp}%
  \BibitemOpen
  \bibfield  {author} {\bibinfo {author} {\bibfnamefont {C.}~\bibnamefont {Dlapa}}, \bibinfo {author} {\bibfnamefont {G.}~\bibnamefont {K\"alin}}, \bibinfo {author} {\bibfnamefont {Z.}~\bibnamefont {Liu}}, \ and\ \bibinfo {author} {\bibfnamefont {R.~A.}\ \bibnamefont {Porto}},\ }\href {\doibase 10.1103/PhysRevLett.128.161104} {\bibfield  {journal} {\bibinfo  {journal} {Phys. Rev. Lett.}\ }\textbf {\bibinfo {volume} {128}},\ \bibinfo {pages} {161104} (\bibinfo {year} {2022}{\natexlab{b}})},\ \Eprint {http://arxiv.org/abs/2112.11296} {arXiv:2112.11296 [hep-th]} \BibitemShut {NoStop}%
\bibitem [{\citenamefont {Damgaard}\ \emph {et~al.}(2023)\citenamefont {Damgaard}, \citenamefont {Hansen}, \citenamefont {Plant{\'e}},\ and\ \citenamefont {Vanhove}}]{Damgaard:2023ttc}%
  \BibitemOpen
  \bibfield  {author} {\bibinfo {author} {\bibfnamefont {P.~H.}\ \bibnamefont {Damgaard}}, \bibinfo {author} {\bibfnamefont {E.~R.}\ \bibnamefont {Hansen}}, \bibinfo {author} {\bibfnamefont {L.}~\bibnamefont {Plant{\'e}}}, \ and\ \bibinfo {author} {\bibfnamefont {P.}~\bibnamefont {Vanhove}},\ }\href {\doibase 10.1007/JHEP09(2023)183} {\bibfield  {journal} {\bibinfo  {journal} {JHEP}\ }\textbf {\bibinfo {volume} {09}},\ \bibinfo {pages} {183} (\bibinfo {year} {2023})},\ \Eprint {http://arxiv.org/abs/2307.04746} {arXiv:2307.04746 [hep-th]} \BibitemShut {NoStop}%
\bibitem [{\citenamefont {Bern}\ \emph {et~al.}(2024{\natexlab{a}})\citenamefont {Bern}, \citenamefont {Herrmann}, \citenamefont {Roiban}, \citenamefont {Ruf}, \citenamefont {Smirnov}, \citenamefont {Smirnov},\ and\ \citenamefont {Zeng}}]{Bern:2023ccb}%
  \BibitemOpen
  \bibfield  {author} {\bibinfo {author} {\bibfnamefont {Z.}~\bibnamefont {Bern}}, \bibinfo {author} {\bibfnamefont {E.}~\bibnamefont {Herrmann}}, \bibinfo {author} {\bibfnamefont {R.}~\bibnamefont {Roiban}}, \bibinfo {author} {\bibfnamefont {M.~S.}\ \bibnamefont {Ruf}}, \bibinfo {author} {\bibfnamefont {A.~V.}\ \bibnamefont {Smirnov}}, \bibinfo {author} {\bibfnamefont {V.~A.}\ \bibnamefont {Smirnov}}, \ and\ \bibinfo {author} {\bibfnamefont {M.}~\bibnamefont {Zeng}},\ }\href {\doibase 10.1103/PhysRevLett.132.251601} {\bibfield  {journal} {\bibinfo  {journal} {Phys. Rev. Lett.}\ }\textbf {\bibinfo {volume} {132}},\ \bibinfo {pages} {251601} (\bibinfo {year} {2024}{\natexlab{a}})},\ \Eprint {http://arxiv.org/abs/2305.08981} {arXiv:2305.08981 [hep-th]} \BibitemShut {NoStop}%
\bibitem [{\citenamefont {Bern}\ \emph {et~al.}(2024{\natexlab{b}})\citenamefont {Bern}, \citenamefont {Herrmann}, \citenamefont {Roiban}, \citenamefont {Ruf}, \citenamefont {Smirnov}, \citenamefont {Smirnov},\ and\ \citenamefont {Zeng}}]{Bern:2024adl}%
  \BibitemOpen
  \bibfield  {author} {\bibinfo {author} {\bibfnamefont {Z.}~\bibnamefont {Bern}}, \bibinfo {author} {\bibfnamefont {E.}~\bibnamefont {Herrmann}}, \bibinfo {author} {\bibfnamefont {R.}~\bibnamefont {Roiban}}, \bibinfo {author} {\bibfnamefont {M.~S.}\ \bibnamefont {Ruf}}, \bibinfo {author} {\bibfnamefont {A.~V.}\ \bibnamefont {Smirnov}}, \bibinfo {author} {\bibfnamefont {V.~A.}\ \bibnamefont {Smirnov}}, \ and\ \bibinfo {author} {\bibfnamefont {M.}~\bibnamefont {Zeng}},\ }\href {\doibase 10.1007/JHEP10(2024)023} {\bibfield  {journal} {\bibinfo  {journal} {JHEP}\ }\textbf {\bibinfo {volume} {10}},\ \bibinfo {pages} {023} (\bibinfo {year} {2024}{\natexlab{b}})},\ \Eprint {http://arxiv.org/abs/2406.01554} {arXiv:2406.01554 [hep-th]} \BibitemShut {NoStop}%
\bibitem [{\citenamefont {Driesse}\ \emph {et~al.}(2024)\citenamefont {Driesse}, \citenamefont {Jakobsen}, \citenamefont {Mogull}, \citenamefont {Plefka}, \citenamefont {Sauer},\ and\ \citenamefont {Usovitsch}}]{Driesse:2024xad}%
  \BibitemOpen
  \bibfield  {author} {\bibinfo {author} {\bibfnamefont {M.}~\bibnamefont {Driesse}}, \bibinfo {author} {\bibfnamefont {G.~U.}\ \bibnamefont {Jakobsen}}, \bibinfo {author} {\bibfnamefont {G.}~\bibnamefont {Mogull}}, \bibinfo {author} {\bibfnamefont {J.}~\bibnamefont {Plefka}}, \bibinfo {author} {\bibfnamefont {B.}~\bibnamefont {Sauer}}, \ and\ \bibinfo {author} {\bibfnamefont {J.}~\bibnamefont {Usovitsch}},\ }\href {\doibase 10.1103/PhysRevLett.132.241402} {\bibfield  {journal} {\bibinfo  {journal} {Phys. Rev. Lett.}\ }\textbf {\bibinfo {volume} {132}},\ \bibinfo {pages} {241402} (\bibinfo {year} {2024})},\ \Eprint {http://arxiv.org/abs/2403.07781} {arXiv:2403.07781 [hep-th]} \BibitemShut {NoStop}%
\bibitem [{\citenamefont {Driesse}\ \emph {et~al.}(2025)\citenamefont {Driesse}, \citenamefont {Jakobsen}, \citenamefont {Klemm}, \citenamefont {Mogull}, \citenamefont {Nega}, \citenamefont {Plefka}, \citenamefont {Sauer},\ and\ \citenamefont {Usovitsch}}]{Driesse:2024feo}%
  \BibitemOpen
  \bibfield  {author} {\bibinfo {author} {\bibfnamefont {M.}~\bibnamefont {Driesse}}, \bibinfo {author} {\bibfnamefont {G.~U.}\ \bibnamefont {Jakobsen}}, \bibinfo {author} {\bibfnamefont {A.}~\bibnamefont {Klemm}}, \bibinfo {author} {\bibfnamefont {G.}~\bibnamefont {Mogull}}, \bibinfo {author} {\bibfnamefont {C.}~\bibnamefont {Nega}}, \bibinfo {author} {\bibfnamefont {J.}~\bibnamefont {Plefka}}, \bibinfo {author} {\bibfnamefont {B.}~\bibnamefont {Sauer}}, \ and\ \bibinfo {author} {\bibfnamefont {J.}~\bibnamefont {Usovitsch}},\ }\href {\doibase 10.1038/s41586-025-08984-2} {\bibfield  {journal} {\bibinfo  {journal} {Nature}\ }\textbf {\bibinfo {volume} {641}},\ \bibinfo {pages} {603} (\bibinfo {year} {2025})},\ \Eprint {http://arxiv.org/abs/2411.11846} {arXiv:2411.11846 [hep-th]} \BibitemShut {NoStop}%
\bibitem [{\citenamefont {Bini}\ and\ \citenamefont {Damour}(2025)}]{Bini:2025vuk}%
  \BibitemOpen
  \bibfield  {author} {\bibinfo {author} {\bibfnamefont {D.}~\bibnamefont {Bini}}\ and\ \bibinfo {author} {\bibfnamefont {T.}~\bibnamefont {Damour}},\ }\href {\doibase 10.1103/8ks7-2blq} {\bibfield  {journal} {\bibinfo  {journal} {Phys. Rev. D}\ }\textbf {\bibinfo {volume} {112}},\ \bibinfo {pages} {044002} (\bibinfo {year} {2025})},\ \Eprint {http://arxiv.org/abs/2504.20204} {arXiv:2504.20204 [hep-th]} \BibitemShut {NoStop}%
\bibitem [{\citenamefont {Bern}\ \emph {et~al.}(2026)\citenamefont {Bern}, \citenamefont {Herrmann}, \citenamefont {Roiban}, \citenamefont {Ruf}, \citenamefont {Smirnov}, \citenamefont {Smirnov},\ and\ \citenamefont {Zeng}}]{Bern:2025zno}%
  \BibitemOpen
  \bibfield  {author} {\bibinfo {author} {\bibfnamefont {Z.}~\bibnamefont {Bern}}, \bibinfo {author} {\bibfnamefont {E.}~\bibnamefont {Herrmann}}, \bibinfo {author} {\bibfnamefont {R.}~\bibnamefont {Roiban}}, \bibinfo {author} {\bibfnamefont {M.~S.}\ \bibnamefont {Ruf}}, \bibinfo {author} {\bibfnamefont {A.~V.}\ \bibnamefont {Smirnov}}, \bibinfo {author} {\bibfnamefont {V.~A.}\ \bibnamefont {Smirnov}}, \ and\ \bibinfo {author} {\bibfnamefont {M.}~\bibnamefont {Zeng}},\ }\href {\doibase 10.1103/jmby-htz9} {\bibfield  {journal} {\bibinfo  {journal} {Phys. Rev. Lett.}\ }\textbf {\bibinfo {volume} {136}},\ \bibinfo {pages} {081401} (\bibinfo {year} {2026})},\ \Eprint {http://arxiv.org/abs/2509.17412} {arXiv:2509.17412 [hep-th]} \BibitemShut {NoStop}%
\bibitem [{\citenamefont {Bern}\ \emph {et~al.}(2025{\natexlab{a}})\citenamefont {Bern}, \citenamefont {Herrmann}, \citenamefont {Roiban}, \citenamefont {Ruf}, \citenamefont {Smirnov}, \citenamefont {Smith},\ and\ \citenamefont {Zeng}}]{Bern:2025wyd}%
  \BibitemOpen
  \bibfield  {author} {\bibinfo {author} {\bibfnamefont {Z.}~\bibnamefont {Bern}}, \bibinfo {author} {\bibfnamefont {E.}~\bibnamefont {Herrmann}}, \bibinfo {author} {\bibfnamefont {R.}~\bibnamefont {Roiban}}, \bibinfo {author} {\bibfnamefont {M.~S.}\ \bibnamefont {Ruf}}, \bibinfo {author} {\bibfnamefont {A.~V.}\ \bibnamefont {Smirnov}}, \bibinfo {author} {\bibfnamefont {S.}~\bibnamefont {Smith}}, \ and\ \bibinfo {author} {\bibfnamefont {M.}~\bibnamefont {Zeng}},\ }\href@noop {} {\  (\bibinfo {year} {2025}{\natexlab{a}})},\ \Eprint {http://arxiv.org/abs/2512.23654} {arXiv:2512.23654 [hep-th]} \BibitemShut {NoStop}%
\bibitem [{\citenamefont {Driesse}\ \emph {et~al.}(2026)\citenamefont {Driesse}, \citenamefont {Jakobsen}, \citenamefont {Mogull}, \citenamefont {Nega}, \citenamefont {Plefka}, \citenamefont {Sauer},\ and\ \citenamefont {Usovitsch}}]{Driesse:2026qiz}%
  \BibitemOpen
  \bibfield  {author} {\bibinfo {author} {\bibfnamefont {M.}~\bibnamefont {Driesse}}, \bibinfo {author} {\bibfnamefont {G.~U.}\ \bibnamefont {Jakobsen}}, \bibinfo {author} {\bibfnamefont {G.}~\bibnamefont {Mogull}}, \bibinfo {author} {\bibfnamefont {C.}~\bibnamefont {Nega}}, \bibinfo {author} {\bibfnamefont {J.}~\bibnamefont {Plefka}}, \bibinfo {author} {\bibfnamefont {B.}~\bibnamefont {Sauer}}, \ and\ \bibinfo {author} {\bibfnamefont {J.}~\bibnamefont {Usovitsch}},\ }\href@noop {} {\  (\bibinfo {year} {2026})},\ \Eprint {http://arxiv.org/abs/2601.16256} {arXiv:2601.16256 [hep-th]} \BibitemShut {NoStop}%
\bibitem [{\citenamefont {Goncharov}(2001)}]{Goncharov:2001iea}%
  \BibitemOpen
  \bibfield  {author} {\bibinfo {author} {\bibfnamefont {A.}~\bibnamefont {Goncharov}},\ }\href@noop {} {\  (\bibinfo {year} {2001})},\ \Eprint {http://arxiv.org/abs/math/0103059} {arXiv:math/0103059} \BibitemShut {NoStop}%
\bibitem [{\citenamefont {Frellesvig}\ \emph {et~al.}(2024)\citenamefont {Frellesvig}, \citenamefont {Morales},\ and\ \citenamefont {Wilhelm}}]{Frellesvig:2023bbf}%
  \BibitemOpen
  \bibfield  {author} {\bibinfo {author} {\bibfnamefont {H.}~\bibnamefont {Frellesvig}}, \bibinfo {author} {\bibfnamefont {R.}~\bibnamefont {Morales}}, \ and\ \bibinfo {author} {\bibfnamefont {M.}~\bibnamefont {Wilhelm}},\ }\href {\doibase 10.1103/PhysRevLett.132.201602} {\bibfield  {journal} {\bibinfo  {journal} {Phys. Rev. Lett.}\ }\textbf {\bibinfo {volume} {132}},\ \bibinfo {pages} {201602} (\bibinfo {year} {2024})},\ \Eprint {http://arxiv.org/abs/2312.11371} {arXiv:2312.11371 [hep-th]} \BibitemShut {NoStop}%
\bibitem [{\citenamefont {Klemm}\ \emph {et~al.}(2024)\citenamefont {Klemm}, \citenamefont {Nega}, \citenamefont {Sauer},\ and\ \citenamefont {Plefka}}]{Klemm:2024wtd}%
  \BibitemOpen
  \bibfield  {author} {\bibinfo {author} {\bibfnamefont {A.}~\bibnamefont {Klemm}}, \bibinfo {author} {\bibfnamefont {C.}~\bibnamefont {Nega}}, \bibinfo {author} {\bibfnamefont {B.}~\bibnamefont {Sauer}}, \ and\ \bibinfo {author} {\bibfnamefont {J.}~\bibnamefont {Plefka}},\ }\href {\doibase 10.1103/PhysRevD.109.124046} {\bibfield  {journal} {\bibinfo  {journal} {Phys. Rev. D}\ }\textbf {\bibinfo {volume} {109}},\ \bibinfo {pages} {124046} (\bibinfo {year} {2024})},\ \Eprint {http://arxiv.org/abs/2401.07899} {arXiv:2401.07899 [hep-th]} \BibitemShut {NoStop}%
\bibitem [{\citenamefont {Chetyrkin}\ and\ \citenamefont {Tkachov}(1981)}]{Chetyrkin:1981qh}%
  \BibitemOpen
  \bibfield  {author} {\bibinfo {author} {\bibfnamefont {K.~G.}\ \bibnamefont {Chetyrkin}}\ and\ \bibinfo {author} {\bibfnamefont {F.~V.}\ \bibnamefont {Tkachov}},\ }\href {\doibase 10.1016/0550-3213(81)90199-1} {\bibfield  {journal} {\bibinfo  {journal} {Nucl. Phys. B}\ }\textbf {\bibinfo {volume} {192}},\ \bibinfo {pages} {159} (\bibinfo {year} {1981})}\BibitemShut {NoStop}%
\bibitem [{\citenamefont {Tkachov}(1981)}]{Tkachov:1981wb}%
  \BibitemOpen
  \bibfield  {author} {\bibinfo {author} {\bibfnamefont {F.~V.}\ \bibnamefont {Tkachov}},\ }\href {\doibase 10.1016/0370-2693(81)90288-4} {\bibfield  {journal} {\bibinfo  {journal} {Phys. Lett. B}\ }\textbf {\bibinfo {volume} {100}},\ \bibinfo {pages} {65} (\bibinfo {year} {1981})}\BibitemShut {NoStop}%
\bibitem [{\citenamefont {Laporta}(2000)}]{Laporta:2000dsw}%
  \BibitemOpen
  \bibfield  {author} {\bibinfo {author} {\bibfnamefont {S.}~\bibnamefont {Laporta}},\ }\href {\doibase 10.1142/S0217751X00002159} {\bibfield  {journal} {\bibinfo  {journal} {Int. J. Mod. Phys. A}\ }\textbf {\bibinfo {volume} {15}},\ \bibinfo {pages} {5087} (\bibinfo {year} {2000})},\ \Eprint {http://arxiv.org/abs/hep-ph/0102033} {arXiv:hep-ph/0102033} \BibitemShut {NoStop}%
\bibitem [{\citenamefont {Henn}(2013)}]{Henn:2013pwa}%
  \BibitemOpen
  \bibfield  {author} {\bibinfo {author} {\bibfnamefont {J.~M.}\ \bibnamefont {Henn}},\ }\href {\doibase 10.1103/PhysRevLett.110.251601} {\bibfield  {journal} {\bibinfo  {journal} {Phys. Rev. Lett.}\ }\textbf {\bibinfo {volume} {110}},\ \bibinfo {pages} {251601} (\bibinfo {year} {2013})},\ \Eprint {http://arxiv.org/abs/1304.1806} {arXiv:1304.1806 [hep-th]} \BibitemShut {NoStop}%
\bibitem [{\citenamefont {Smirnov}\ and\ \citenamefont {Smirnov}(2020)}]{Smirnov:2020quc}%
  \BibitemOpen
  \bibfield  {author} {\bibinfo {author} {\bibfnamefont {A.~V.}\ \bibnamefont {Smirnov}}\ and\ \bibinfo {author} {\bibfnamefont {V.~A.}\ \bibnamefont {Smirnov}},\ }\href {\doibase 10.1016/j.nuclphysb.2020.115213} {\bibfield  {journal} {\bibinfo  {journal} {Nucl. Phys. B}\ }\textbf {\bibinfo {volume} {960}},\ \bibinfo {pages} {115213} (\bibinfo {year} {2020})},\ \Eprint {http://arxiv.org/abs/2002.08042} {arXiv:2002.08042 [hep-ph]} \BibitemShut {NoStop}%
\bibitem [{\citenamefont {Usovitsch}(2020)}]{Usovitsch:2020jrk}%
  \BibitemOpen
  \bibfield  {author} {\bibinfo {author} {\bibfnamefont {J.}~\bibnamefont {Usovitsch}},\ }\href@noop {} {\  (\bibinfo {year} {2020})},\ \Eprint {http://arxiv.org/abs/2002.08173} {arXiv:2002.08173 [hep-ph]} \BibitemShut {NoStop}%
\bibitem [{\citenamefont {Smirnov}(2008)}]{Smirnov:2008iw}%
  \BibitemOpen
  \bibfield  {author} {\bibinfo {author} {\bibfnamefont {A.~V.}\ \bibnamefont {Smirnov}},\ }\href {\doibase 10.1088/1126-6708/2008/10/107} {\bibfield  {journal} {\bibinfo  {journal} {JHEP}\ }\textbf {\bibinfo {volume} {10}},\ \bibinfo {pages} {107} (\bibinfo {year} {2008})},\ \Eprint {http://arxiv.org/abs/0807.3243} {arXiv:0807.3243 [hep-ph]} \BibitemShut {NoStop}%
\bibitem [{\citenamefont {Maierh\"ofer}\ \emph {et~al.}(2018)\citenamefont {Maierh\"ofer}, \citenamefont {Usovitsch},\ and\ \citenamefont {Uwer}}]{Maierhofer:2017gsa}%
  \BibitemOpen
  \bibfield  {author} {\bibinfo {author} {\bibfnamefont {P.}~\bibnamefont {Maierh\"ofer}}, \bibinfo {author} {\bibfnamefont {J.}~\bibnamefont {Usovitsch}}, \ and\ \bibinfo {author} {\bibfnamefont {P.}~\bibnamefont {Uwer}},\ }\href {\doibase 10.1016/j.cpc.2018.04.012} {\bibfield  {journal} {\bibinfo  {journal} {Comput. Phys. Commun.}\ }\textbf {\bibinfo {volume} {230}},\ \bibinfo {pages} {99} (\bibinfo {year} {2018})},\ \Eprint {http://arxiv.org/abs/1705.05610} {arXiv:1705.05610 [hep-ph]} \BibitemShut {NoStop}%
\bibitem [{\citenamefont {Guan}\ \emph {et~al.}(2025)\citenamefont {Guan}, \citenamefont {Liu}, \citenamefont {Ma},\ and\ \citenamefont {Wu}}]{Guan:2024byi}%
  \BibitemOpen
  \bibfield  {author} {\bibinfo {author} {\bibfnamefont {X.}~\bibnamefont {Guan}}, \bibinfo {author} {\bibfnamefont {X.}~\bibnamefont {Liu}}, \bibinfo {author} {\bibfnamefont {Y.-Q.}\ \bibnamefont {Ma}}, \ and\ \bibinfo {author} {\bibfnamefont {W.-H.}\ \bibnamefont {Wu}},\ }\href {\doibase 10.1016/j.cpc.2025.109538} {\bibfield  {journal} {\bibinfo  {journal} {Comput. Phys. Commun.}\ }\textbf {\bibinfo {volume} {310}},\ \bibinfo {pages} {109538} (\bibinfo {year} {2025})},\ \Eprint {http://arxiv.org/abs/2405.14621} {arXiv:2405.14621 [hep-ph]} \BibitemShut {NoStop}%
\bibitem [{\citenamefont {Smirnov}\ and\ \citenamefont {Zeng}(2025)}]{FIRE7}%
  \BibitemOpen
  \bibfield  {author} {\bibinfo {author} {\bibfnamefont {A.~V.}\ \bibnamefont {Smirnov}}\ and\ \bibinfo {author} {\bibfnamefont {M.}~\bibnamefont {Zeng}},\ }\href@noop {} {\  (\bibinfo {year} {2025})},\ \Eprint {http://arxiv.org/abs/2510.07150} {arXiv:2510.07150 [hep-ph]} \BibitemShut {NoStop}%
\bibitem [{\citenamefont {Lange}\ \emph {et~al.}(2025)\citenamefont {Lange}, \citenamefont {Usovitsch},\ and\ \citenamefont {Wu}}]{Lange:2025fba}%
  \BibitemOpen
  \bibfield  {author} {\bibinfo {author} {\bibfnamefont {F.}~\bibnamefont {Lange}}, \bibinfo {author} {\bibfnamefont {J.}~\bibnamefont {Usovitsch}}, \ and\ \bibinfo {author} {\bibfnamefont {Z.}~\bibnamefont {Wu}},\ }\href@noop {} {\  (\bibinfo {year} {2025})},\ \Eprint {http://arxiv.org/abs/2505.20197} {arXiv:2505.20197 [hep-ph]} \BibitemShut {NoStop}%
\bibitem [{\citenamefont {Lee}(2014)}]{Lee:2013mka}%
  \BibitemOpen
  \bibfield  {author} {\bibinfo {author} {\bibfnamefont {R.~N.}\ \bibnamefont {Lee}},\ }\href {\doibase 10.1088/1742-6596/523/1/012059} {\bibfield  {journal} {\bibinfo  {journal} {J. Phys. Conf. Ser.}\ }\textbf {\bibinfo {volume} {523}},\ \bibinfo {pages} {012059} (\bibinfo {year} {2014})},\ \Eprint {http://arxiv.org/abs/1310.1145} {arXiv:1310.1145 [hep-ph]} \BibitemShut {NoStop}%
\bibitem [{\citenamefont {Kotikov}(1991)}]{Kotikov:1990kg}%
  \BibitemOpen
  \bibfield  {author} {\bibinfo {author} {\bibfnamefont {A.~V.}\ \bibnamefont {Kotikov}},\ }\href {\doibase 10.1016/0370-2693(91)90413-K} {\bibfield  {journal} {\bibinfo  {journal} {Phys. Lett. B}\ }\textbf {\bibinfo {volume} {254}},\ \bibinfo {pages} {158} (\bibinfo {year} {1991})}\BibitemShut {NoStop}%
\bibitem [{\citenamefont {Bern}\ \emph {et~al.}(1994{\natexlab{a}})\citenamefont {Bern}, \citenamefont {Dixon},\ and\ \citenamefont {Kosower}}]{Bern:1993kr}%
  \BibitemOpen
  \bibfield  {author} {\bibinfo {author} {\bibfnamefont {Z.}~\bibnamefont {Bern}}, \bibinfo {author} {\bibfnamefont {L.~J.}\ \bibnamefont {Dixon}}, \ and\ \bibinfo {author} {\bibfnamefont {D.~A.}\ \bibnamefont {Kosower}},\ }\href {\doibase 10.1016/0550-3213(94)90398-0} {\bibfield  {journal} {\bibinfo  {journal} {Nucl. Phys. B}\ }\textbf {\bibinfo {volume} {412}},\ \bibinfo {pages} {751} (\bibinfo {year} {1994}{\natexlab{a}})},\ \Eprint {http://arxiv.org/abs/hep-ph/9306240} {arXiv:hep-ph/9306240} \BibitemShut {NoStop}%
\bibitem [{\citenamefont {Remiddi}(1997)}]{Remiddi:1997ny}%
  \BibitemOpen
  \bibfield  {author} {\bibinfo {author} {\bibfnamefont {E.}~\bibnamefont {Remiddi}},\ }\href {\doibase 10.1007/BF03185566} {\bibfield  {journal} {\bibinfo  {journal} {Nuovo Cim. A}\ }\textbf {\bibinfo {volume} {110}},\ \bibinfo {pages} {1435} (\bibinfo {year} {1997})},\ \Eprint {http://arxiv.org/abs/hep-th/9711188} {arXiv:hep-th/9711188} \BibitemShut {NoStop}%
\bibitem [{\citenamefont {Gehrmann}\ and\ \citenamefont {Remiddi}(2000)}]{Gehrmann:1999as}%
  \BibitemOpen
  \bibfield  {author} {\bibinfo {author} {\bibfnamefont {T.}~\bibnamefont {Gehrmann}}\ and\ \bibinfo {author} {\bibfnamefont {E.}~\bibnamefont {Remiddi}},\ }\href {\doibase 10.1016/S0550-3213(00)00223-6} {\bibfield  {journal} {\bibinfo  {journal} {Nucl. Phys. B}\ }\textbf {\bibinfo {volume} {580}},\ \bibinfo {pages} {485} (\bibinfo {year} {2000})},\ \Eprint {http://arxiv.org/abs/hep-ph/9912329} {arXiv:hep-ph/9912329} \BibitemShut {NoStop}%
\bibitem [{\citenamefont {Caron-Huot}\ and\ \citenamefont {Wilhelm}(2016)}]{Caron-Huot:2016cwu}%
  \BibitemOpen
  \bibfield  {author} {\bibinfo {author} {\bibfnamefont {S.}~\bibnamefont {Caron-Huot}}\ and\ \bibinfo {author} {\bibfnamefont {M.}~\bibnamefont {Wilhelm}},\ }\href {\doibase 10.1007/JHEP12(2016)010} {\bibfield  {journal} {\bibinfo  {journal} {JHEP}\ }\textbf {\bibinfo {volume} {12}},\ \bibinfo {pages} {010} (\bibinfo {year} {2016})},\ \Eprint {http://arxiv.org/abs/1607.06448} {arXiv:1607.06448 [hep-th]} \BibitemShut {NoStop}%
\bibitem [{\citenamefont {Herzog}\ \emph {et~al.}(2017)\citenamefont {Herzog}, \citenamefont {Ruijl}, \citenamefont {Ueda}, \citenamefont {Vermaseren},\ and\ \citenamefont {Vogt}}]{Herzog:2017ohr}%
  \BibitemOpen
  \bibfield  {author} {\bibinfo {author} {\bibfnamefont {F.}~\bibnamefont {Herzog}}, \bibinfo {author} {\bibfnamefont {B.}~\bibnamefont {Ruijl}}, \bibinfo {author} {\bibfnamefont {T.}~\bibnamefont {Ueda}}, \bibinfo {author} {\bibfnamefont {J.~A.~M.}\ \bibnamefont {Vermaseren}}, \ and\ \bibinfo {author} {\bibfnamefont {A.}~\bibnamefont {Vogt}},\ }\href {\doibase 10.1007/JHEP02(2017)090} {\bibfield  {journal} {\bibinfo  {journal} {JHEP}\ }\textbf {\bibinfo {volume} {02}},\ \bibinfo {pages} {090} (\bibinfo {year} {2017})},\ \Eprint {http://arxiv.org/abs/1701.01404} {arXiv:1701.01404 [hep-ph]} \BibitemShut {NoStop}%
\bibitem [{\citenamefont {Bern}\ \emph {et~al.}(2018)\citenamefont {Bern}, \citenamefont {Carrasco}, \citenamefont {Chen}, \citenamefont {Edison}, \citenamefont {Johansson}, \citenamefont {Parra-Martinez}, \citenamefont {Roiban},\ and\ \citenamefont {Zeng}}]{Bern:2018jmv}%
  \BibitemOpen
  \bibfield  {author} {\bibinfo {author} {\bibfnamefont {Z.}~\bibnamefont {Bern}}, \bibinfo {author} {\bibfnamefont {J.~J.}\ \bibnamefont {Carrasco}}, \bibinfo {author} {\bibfnamefont {W.-M.}\ \bibnamefont {Chen}}, \bibinfo {author} {\bibfnamefont {A.}~\bibnamefont {Edison}}, \bibinfo {author} {\bibfnamefont {H.}~\bibnamefont {Johansson}}, \bibinfo {author} {\bibfnamefont {J.}~\bibnamefont {Parra-Martinez}}, \bibinfo {author} {\bibfnamefont {R.}~\bibnamefont {Roiban}}, \ and\ \bibinfo {author} {\bibfnamefont {M.}~\bibnamefont {Zeng}},\ }\href {\doibase 10.1103/PhysRevD.98.086021} {\bibfield  {journal} {\bibinfo  {journal} {Phys. Rev. D}\ }\textbf {\bibinfo {volume} {98}},\ \bibinfo {pages} {086021} (\bibinfo {year} {2018})},\ \Eprint {http://arxiv.org/abs/1804.09311} {arXiv:1804.09311 [hep-th]} \BibitemShut {NoStop}%
\bibitem [{\citenamefont {Bern}\ \emph {et~al.}(2020)\citenamefont {Bern}, \citenamefont {Parra-Martinez},\ and\ \citenamefont {Sawyer}}]{Bern:2020ikv}%
  \BibitemOpen
  \bibfield  {author} {\bibinfo {author} {\bibfnamefont {Z.}~\bibnamefont {Bern}}, \bibinfo {author} {\bibfnamefont {J.}~\bibnamefont {Parra-Martinez}}, \ and\ \bibinfo {author} {\bibfnamefont {E.}~\bibnamefont {Sawyer}},\ }\href {\doibase 10.1007/JHEP10(2020)211} {\bibfield  {journal} {\bibinfo  {journal} {JHEP}\ }\textbf {\bibinfo {volume} {10}},\ \bibinfo {pages} {211} (\bibinfo {year} {2020})},\ \Eprint {http://arxiv.org/abs/2005.12917} {arXiv:2005.12917 [hep-ph]} \BibitemShut {NoStop}%
\bibitem [{\citenamefont {Weinberg}(1965)}]{Weinberg:1965nx}%
  \BibitemOpen
  \bibfield  {author} {\bibinfo {author} {\bibfnamefont {S.}~\bibnamefont {Weinberg}},\ }\href {\doibase 10.1103/PhysRev.140.B516} {\bibfield  {journal} {\bibinfo  {journal} {Phys. Rev.}\ }\textbf {\bibinfo {volume} {140}},\ \bibinfo {pages} {B516} (\bibinfo {year} {1965})}\BibitemShut {NoStop}%
\bibitem [{\citenamefont {Naculich}\ and\ \citenamefont {Schnitzer}(2011)}]{Naculich:2011ry}%
  \BibitemOpen
  \bibfield  {author} {\bibinfo {author} {\bibfnamefont {S.~G.}\ \bibnamefont {Naculich}}\ and\ \bibinfo {author} {\bibfnamefont {H.~J.}\ \bibnamefont {Schnitzer}},\ }\href {\doibase 10.1007/JHEP05(2011)087} {\bibfield  {journal} {\bibinfo  {journal} {JHEP}\ }\textbf {\bibinfo {volume} {05}},\ \bibinfo {pages} {087} (\bibinfo {year} {2011})},\ \Eprint {http://arxiv.org/abs/1101.1524} {arXiv:1101.1524 [hep-th]} \BibitemShut {NoStop}%
\bibitem [{\citenamefont {Akhoury}\ \emph {et~al.}(2011)\citenamefont {Akhoury}, \citenamefont {Saotome},\ and\ \citenamefont {Sterman}}]{Akhoury:2011kq}%
  \BibitemOpen
  \bibfield  {author} {\bibinfo {author} {\bibfnamefont {R.}~\bibnamefont {Akhoury}}, \bibinfo {author} {\bibfnamefont {R.}~\bibnamefont {Saotome}}, \ and\ \bibinfo {author} {\bibfnamefont {G.}~\bibnamefont {Sterman}},\ }\href {\doibase 10.1103/PhysRevD.84.104040} {\bibfield  {journal} {\bibinfo  {journal} {Phys. Rev. D}\ }\textbf {\bibinfo {volume} {84}},\ \bibinfo {pages} {104040} (\bibinfo {year} {2011})},\ \Eprint {http://arxiv.org/abs/1109.0270} {arXiv:1109.0270 [hep-th]} \BibitemShut {NoStop}%
\bibitem [{\citenamefont {Anastasiou}\ and\ \citenamefont {Sterman}(2019)}]{Anastasiou:2018rib}%
  \BibitemOpen
  \bibfield  {author} {\bibinfo {author} {\bibfnamefont {C.}~\bibnamefont {Anastasiou}}\ and\ \bibinfo {author} {\bibfnamefont {G.}~\bibnamefont {Sterman}},\ }\href {\doibase 10.1007/JHEP07(2019)056} {\bibfield  {journal} {\bibinfo  {journal} {JHEP}\ }\textbf {\bibinfo {volume} {07}},\ \bibinfo {pages} {056} (\bibinfo {year} {2019})},\ \Eprint {http://arxiv.org/abs/1812.03753} {arXiv:1812.03753 [hep-ph]} \BibitemShut {NoStop}%
\bibitem [{\citenamefont {Anastasiou}\ \emph {et~al.}(2021)\citenamefont {Anastasiou}, \citenamefont {Haindl}, \citenamefont {Sterman}, \citenamefont {Yang},\ and\ \citenamefont {Zeng}}]{Anastasiou:2020sdt}%
  \BibitemOpen
  \bibfield  {author} {\bibinfo {author} {\bibfnamefont {C.}~\bibnamefont {Anastasiou}}, \bibinfo {author} {\bibfnamefont {R.}~\bibnamefont {Haindl}}, \bibinfo {author} {\bibfnamefont {G.}~\bibnamefont {Sterman}}, \bibinfo {author} {\bibfnamefont {Z.}~\bibnamefont {Yang}}, \ and\ \bibinfo {author} {\bibfnamefont {M.}~\bibnamefont {Zeng}},\ }\href {\doibase 10.1007/JHEP04(2021)222} {\bibfield  {journal} {\bibinfo  {journal} {JHEP}\ }\textbf {\bibinfo {volume} {04}},\ \bibinfo {pages} {222} (\bibinfo {year} {2021})},\ \Eprint {http://arxiv.org/abs/2008.12293} {arXiv:2008.12293 [hep-ph]} \BibitemShut {NoStop}%
\bibitem [{\citenamefont {Heissenberg}(2021)}]{Heissenberg:2021tzo}%
  \BibitemOpen
  \bibfield  {author} {\bibinfo {author} {\bibfnamefont {C.}~\bibnamefont {Heissenberg}},\ }\href {\doibase 10.1103/PhysRevD.104.046016} {\bibfield  {journal} {\bibinfo  {journal} {Phys. Rev. D}\ }\textbf {\bibinfo {volume} {104}},\ \bibinfo {pages} {046016} (\bibinfo {year} {2021})},\ \Eprint {http://arxiv.org/abs/2105.04594} {arXiv:2105.04594 [hep-th]} \BibitemShut {NoStop}%
\bibitem [{\citenamefont {Damour}(2020)}]{Damour:2020tta}%
  \BibitemOpen
  \bibfield  {author} {\bibinfo {author} {\bibfnamefont {T.}~\bibnamefont {Damour}},\ }\href {\doibase 10.1103/PhysRevD.102.124008} {\bibfield  {journal} {\bibinfo  {journal} {Phys. Rev. D}\ }\textbf {\bibinfo {volume} {102}},\ \bibinfo {pages} {124008} (\bibinfo {year} {2020})},\ \Eprint {http://arxiv.org/abs/2010.01641} {arXiv:2010.01641 [gr-qc]} \BibitemShut {NoStop}%
\bibitem [{\citenamefont {Herrmann}\ \emph {et~al.}(2021{\natexlab{a}})\citenamefont {Herrmann}, \citenamefont {Parra-Martinez}, \citenamefont {Ruf},\ and\ \citenamefont {Zeng}}]{Herrmann:2021tct}%
  \BibitemOpen
  \bibfield  {author} {\bibinfo {author} {\bibfnamefont {E.}~\bibnamefont {Herrmann}}, \bibinfo {author} {\bibfnamefont {J.}~\bibnamefont {Parra-Martinez}}, \bibinfo {author} {\bibfnamefont {M.~S.}\ \bibnamefont {Ruf}}, \ and\ \bibinfo {author} {\bibfnamefont {M.}~\bibnamefont {Zeng}},\ }\href {\doibase 10.1007/JHEP10(2021)148} {\bibfield  {journal} {\bibinfo  {journal} {JHEP}\ }\textbf {\bibinfo {volume} {10}},\ \bibinfo {pages} {148} (\bibinfo {year} {2021}{\natexlab{a}})},\ \Eprint {http://arxiv.org/abs/2104.03957} {arXiv:2104.03957 [hep-th]} \BibitemShut {NoStop}%
\bibitem [{\citenamefont {Dlapa}\ \emph {et~al.}(2023)\citenamefont {Dlapa}, \citenamefont {K\"alin}, \citenamefont {Liu}, \citenamefont {Neef},\ and\ \citenamefont {Porto}}]{Dlapa:2022lmu}%
  \BibitemOpen
  \bibfield  {author} {\bibinfo {author} {\bibfnamefont {C.}~\bibnamefont {Dlapa}}, \bibinfo {author} {\bibfnamefont {G.}~\bibnamefont {K\"alin}}, \bibinfo {author} {\bibfnamefont {Z.}~\bibnamefont {Liu}}, \bibinfo {author} {\bibfnamefont {J.}~\bibnamefont {Neef}}, \ and\ \bibinfo {author} {\bibfnamefont {R.~A.}\ \bibnamefont {Porto}},\ }\href {\doibase 10.1103/PhysRevLett.130.101401} {\bibfield  {journal} {\bibinfo  {journal} {Phys. Rev. Lett.}\ }\textbf {\bibinfo {volume} {130}},\ \bibinfo {pages} {101401} (\bibinfo {year} {2023})},\ \Eprint {http://arxiv.org/abs/2210.05541} {arXiv:2210.05541 [hep-th]} \BibitemShut {NoStop}%
\bibitem [{\citenamefont {Kosower}\ \emph {et~al.}(2019)\citenamefont {Kosower}, \citenamefont {Maybee},\ and\ \citenamefont {O'Connell}}]{Kosower:2018adc}%
  \BibitemOpen
  \bibfield  {author} {\bibinfo {author} {\bibfnamefont {D.~A.}\ \bibnamefont {Kosower}}, \bibinfo {author} {\bibfnamefont {B.}~\bibnamefont {Maybee}}, \ and\ \bibinfo {author} {\bibfnamefont {D.}~\bibnamefont {O'Connell}},\ }\href {\doibase 10.1007/JHEP02(2019)137} {\bibfield  {journal} {\bibinfo  {journal} {JHEP}\ }\textbf {\bibinfo {volume} {02}},\ \bibinfo {pages} {137} (\bibinfo {year} {2019})},\ \Eprint {http://arxiv.org/abs/1811.10950} {arXiv:1811.10950 [hep-th]} \BibitemShut {NoStop}%
\bibitem [{\citenamefont {Brandhuber}\ \emph {et~al.}(2025)\citenamefont {Brandhuber}, \citenamefont {Brown}, \citenamefont {Pichini}, \citenamefont {Travaglini},\ and\ \citenamefont {Vives~Matasan}}]{Brandhuber:2025igz}%
  \BibitemOpen
  \bibfield  {author} {\bibinfo {author} {\bibfnamefont {A.}~\bibnamefont {Brandhuber}}, \bibinfo {author} {\bibfnamefont {G.~R.}\ \bibnamefont {Brown}}, \bibinfo {author} {\bibfnamefont {P.}~\bibnamefont {Pichini}}, \bibinfo {author} {\bibfnamefont {G.}~\bibnamefont {Travaglini}}, \ and\ \bibinfo {author} {\bibfnamefont {P.}~\bibnamefont {Vives~Matasan}},\ }\href@noop {} {\  (\bibinfo {year} {2025})},\ \Eprint {http://arxiv.org/abs/2512.05017} {arXiv:2512.05017 [hep-th]} \BibitemShut {NoStop}%
\bibitem [{\citenamefont {Kim}\ \emph {et~al.}(2025)\citenamefont {Kim}, \citenamefont {Patil}, \citenamefont {Scheopner},\ and\ \citenamefont {Steinhoff}}]{Kim:2025gis}%
  \BibitemOpen
  \bibfield  {author} {\bibinfo {author} {\bibfnamefont {J.-W.}\ \bibnamefont {Kim}}, \bibinfo {author} {\bibfnamefont {R.}~\bibnamefont {Patil}}, \bibinfo {author} {\bibfnamefont {T.}~\bibnamefont {Scheopner}}, \ and\ \bibinfo {author} {\bibfnamefont {J.}~\bibnamefont {Steinhoff}},\ }\href@noop {} {\  (\bibinfo {year} {2025})},\ \Eprint {http://arxiv.org/abs/2511.05649} {arXiv:2511.05649 [hep-th]} \BibitemShut {NoStop}%
\bibitem [{\citenamefont {Bern}\ \emph {et~al.}(1994{\natexlab{b}})\citenamefont {Bern}, \citenamefont {Dixon}, \citenamefont {Dunbar},\ and\ \citenamefont {Kosower}}]{Bern:1994zx}%
  \BibitemOpen
  \bibfield  {author} {\bibinfo {author} {\bibfnamefont {Z.}~\bibnamefont {Bern}}, \bibinfo {author} {\bibfnamefont {L.~J.}\ \bibnamefont {Dixon}}, \bibinfo {author} {\bibfnamefont {D.~C.}\ \bibnamefont {Dunbar}}, \ and\ \bibinfo {author} {\bibfnamefont {D.~A.}\ \bibnamefont {Kosower}},\ }\href {\doibase 10.1016/0550-3213(94)90179-1} {\bibfield  {journal} {\bibinfo  {journal} {Nucl. Phys. B}\ }\textbf {\bibinfo {volume} {425}},\ \bibinfo {pages} {217} (\bibinfo {year} {1994}{\natexlab{b}})},\ \Eprint {http://arxiv.org/abs/hep-ph/9403226} {arXiv:hep-ph/9403226} \BibitemShut {NoStop}%
\bibitem [{\citenamefont {Bern}\ \emph {et~al.}(1995)\citenamefont {Bern}, \citenamefont {Dixon}, \citenamefont {Dunbar},\ and\ \citenamefont {Kosower}}]{Bern:1994cg}%
  \BibitemOpen
  \bibfield  {author} {\bibinfo {author} {\bibfnamefont {Z.}~\bibnamefont {Bern}}, \bibinfo {author} {\bibfnamefont {L.~J.}\ \bibnamefont {Dixon}}, \bibinfo {author} {\bibfnamefont {D.~C.}\ \bibnamefont {Dunbar}}, \ and\ \bibinfo {author} {\bibfnamefont {D.~A.}\ \bibnamefont {Kosower}},\ }\href {\doibase 10.1016/0550-3213(94)00488-Z} {\bibfield  {journal} {\bibinfo  {journal} {Nucl. Phys. B}\ }\textbf {\bibinfo {volume} {435}},\ \bibinfo {pages} {59} (\bibinfo {year} {1995})},\ \Eprint {http://arxiv.org/abs/hep-ph/9409265} {arXiv:hep-ph/9409265} \BibitemShut {NoStop}%
\bibitem [{\citenamefont {Bern}\ \emph {et~al.}(1998)\citenamefont {Bern}, \citenamefont {Dixon},\ and\ \citenamefont {Kosower}}]{Bern:1997sc}%
  \BibitemOpen
  \bibfield  {author} {\bibinfo {author} {\bibfnamefont {Z.}~\bibnamefont {Bern}}, \bibinfo {author} {\bibfnamefont {L.~J.}\ \bibnamefont {Dixon}}, \ and\ \bibinfo {author} {\bibfnamefont {D.~A.}\ \bibnamefont {Kosower}},\ }\href {\doibase 10.1016/S0550-3213(97)00703-7} {\bibfield  {journal} {\bibinfo  {journal} {Nucl. Phys. B}\ }\textbf {\bibinfo {volume} {513}},\ \bibinfo {pages} {3} (\bibinfo {year} {1998})},\ \Eprint {http://arxiv.org/abs/hep-ph/9708239} {arXiv:hep-ph/9708239} \BibitemShut {NoStop}%
\bibitem [{\citenamefont {Britto}\ \emph {et~al.}(2005)\citenamefont {Britto}, \citenamefont {Cachazo},\ and\ \citenamefont {Feng}}]{Britto:2004nc}%
  \BibitemOpen
  \bibfield  {author} {\bibinfo {author} {\bibfnamefont {R.}~\bibnamefont {Britto}}, \bibinfo {author} {\bibfnamefont {F.}~\bibnamefont {Cachazo}}, \ and\ \bibinfo {author} {\bibfnamefont {B.}~\bibnamefont {Feng}},\ }\href {\doibase 10.1016/j.nuclphysb.2005.07.014} {\bibfield  {journal} {\bibinfo  {journal} {Nucl. Phys. B}\ }\textbf {\bibinfo {volume} {725}},\ \bibinfo {pages} {275} (\bibinfo {year} {2005})},\ \Eprint {http://arxiv.org/abs/hep-th/0412103} {arXiv:hep-th/0412103} \BibitemShut {NoStop}%
\bibitem [{\citenamefont {Bern}\ \emph {et~al.}(2004)\citenamefont {Bern}, \citenamefont {Dixon},\ and\ \citenamefont {Kosower}}]{Bern:2004cz}%
  \BibitemOpen
  \bibfield  {author} {\bibinfo {author} {\bibfnamefont {Z.}~\bibnamefont {Bern}}, \bibinfo {author} {\bibfnamefont {L.~J.}\ \bibnamefont {Dixon}}, \ and\ \bibinfo {author} {\bibfnamefont {D.~A.}\ \bibnamefont {Kosower}},\ }\href {\doibase 10.1088/1126-6708/2004/08/012} {\bibfield  {journal} {\bibinfo  {journal} {JHEP}\ }\textbf {\bibinfo {volume} {08}},\ \bibinfo {pages} {012} (\bibinfo {year} {2004})},\ \Eprint {http://arxiv.org/abs/hep-ph/0404293} {arXiv:hep-ph/0404293} \BibitemShut {NoStop}%
\bibitem [{\citenamefont {Bern}\ \emph {et~al.}(2007)\citenamefont {Bern}, \citenamefont {Carrasco}, \citenamefont {Johansson},\ and\ \citenamefont {Kosower}}]{Bern:2007ct}%
  \BibitemOpen
  \bibfield  {author} {\bibinfo {author} {\bibfnamefont {Z.}~\bibnamefont {Bern}}, \bibinfo {author} {\bibfnamefont {J.~J.~M.}\ \bibnamefont {Carrasco}}, \bibinfo {author} {\bibfnamefont {H.}~\bibnamefont {Johansson}}, \ and\ \bibinfo {author} {\bibfnamefont {D.~A.}\ \bibnamefont {Kosower}},\ }\href {\doibase 10.1103/PhysRevD.76.125020} {\bibfield  {journal} {\bibinfo  {journal} {Phys. Rev. D}\ }\textbf {\bibinfo {volume} {76}},\ \bibinfo {pages} {125020} (\bibinfo {year} {2007})},\ \Eprint {http://arxiv.org/abs/0705.1864} {arXiv:0705.1864 [hep-th]} \BibitemShut {NoStop}%
\bibitem [{\citenamefont {Kawai}\ \emph {et~al.}(1986)\citenamefont {Kawai}, \citenamefont {Lewellen},\ and\ \citenamefont {Tye}}]{Kawai:1985xq}%
  \BibitemOpen
  \bibfield  {author} {\bibinfo {author} {\bibfnamefont {H.}~\bibnamefont {Kawai}}, \bibinfo {author} {\bibfnamefont {D.~C.}\ \bibnamefont {Lewellen}}, \ and\ \bibinfo {author} {\bibfnamefont {S.~H.~H.}\ \bibnamefont {Tye}},\ }\href {\doibase 10.1016/0550-3213(86)90362-7} {\bibfield  {journal} {\bibinfo  {journal} {Nucl. Phys. B}\ }\textbf {\bibinfo {volume} {269}},\ \bibinfo {pages} {1} (\bibinfo {year} {1986})}\BibitemShut {NoStop}%
\bibitem [{\citenamefont {Bern}\ \emph {et~al.}(2008)\citenamefont {Bern}, \citenamefont {Carrasco},\ and\ \citenamefont {Johansson}}]{Bern:2008qj}%
  \BibitemOpen
  \bibfield  {author} {\bibinfo {author} {\bibfnamefont {Z.}~\bibnamefont {Bern}}, \bibinfo {author} {\bibfnamefont {J.~J.~M.}\ \bibnamefont {Carrasco}}, \ and\ \bibinfo {author} {\bibfnamefont {H.}~\bibnamefont {Johansson}},\ }\href {\doibase 10.1103/PhysRevD.78.085011} {\bibfield  {journal} {\bibinfo  {journal} {Phys. Rev. D}\ }\textbf {\bibinfo {volume} {78}},\ \bibinfo {pages} {085011} (\bibinfo {year} {2008})},\ \Eprint {http://arxiv.org/abs/0805.3993} {arXiv:0805.3993 [hep-ph]} \BibitemShut {NoStop}%
\bibitem [{\citenamefont {Bern}\ \emph {et~al.}(2010)\citenamefont {Bern}, \citenamefont {Carrasco},\ and\ \citenamefont {Johansson}}]{Bern:2010ue}%
  \BibitemOpen
  \bibfield  {author} {\bibinfo {author} {\bibfnamefont {Z.}~\bibnamefont {Bern}}, \bibinfo {author} {\bibfnamefont {J.~J.~M.}\ \bibnamefont {Carrasco}}, \ and\ \bibinfo {author} {\bibfnamefont {H.}~\bibnamefont {Johansson}},\ }\href {\doibase 10.1103/PhysRevLett.105.061602} {\bibfield  {journal} {\bibinfo  {journal} {Phys. Rev. Lett.}\ }\textbf {\bibinfo {volume} {105}},\ \bibinfo {pages} {061602} (\bibinfo {year} {2010})},\ \Eprint {http://arxiv.org/abs/1004.0476} {arXiv:1004.0476 [hep-th]} \BibitemShut {NoStop}%
\bibitem [{\citenamefont {Bern}\ \emph {et~al.}(2024{\natexlab{c}})\citenamefont {Bern}, \citenamefont {Carrasco}, \citenamefont {Chiodaroli}, \citenamefont {Johansson},\ and\ \citenamefont {Roiban}}]{Bern:2019prr}%
  \BibitemOpen
  \bibfield  {author} {\bibinfo {author} {\bibfnamefont {Z.}~\bibnamefont {Bern}}, \bibinfo {author} {\bibfnamefont {J.~J.}\ \bibnamefont {Carrasco}}, \bibinfo {author} {\bibfnamefont {M.}~\bibnamefont {Chiodaroli}}, \bibinfo {author} {\bibfnamefont {H.}~\bibnamefont {Johansson}}, \ and\ \bibinfo {author} {\bibfnamefont {R.}~\bibnamefont {Roiban}},\ }\href {\doibase 10.1088/1751-8121/ad5fd0} {\bibfield  {journal} {\bibinfo  {journal} {J. Phys. A}\ }\textbf {\bibinfo {volume} {57}},\ \bibinfo {pages} {333002} (\bibinfo {year} {2024}{\natexlab{c}})},\ \Eprint {http://arxiv.org/abs/1909.01358} {arXiv:1909.01358 [hep-th]} \BibitemShut {NoStop}%
\bibitem [{\citenamefont {Beneke}\ and\ \citenamefont {Smirnov}(1998)}]{Beneke:1997zp}%
  \BibitemOpen
  \bibfield  {author} {\bibinfo {author} {\bibfnamefont {M.}~\bibnamefont {Beneke}}\ and\ \bibinfo {author} {\bibfnamefont {V.~A.}\ \bibnamefont {Smirnov}},\ }\href {\doibase 10.1016/S0550-3213(98)00138-2} {\bibfield  {journal} {\bibinfo  {journal} {Nucl. Phys. B}\ }\textbf {\bibinfo {volume} {522}},\ \bibinfo {pages} {321} (\bibinfo {year} {1998})},\ \Eprint {http://arxiv.org/abs/hep-ph/9711391} {arXiv:hep-ph/9711391} \BibitemShut {NoStop}%
\bibitem [{\citenamefont {Bern}\ \emph {et~al.}(2025{\natexlab{b}})\citenamefont {Bern}, \citenamefont {Herrmann}, \citenamefont {Roiban}, \citenamefont {Ruf},\ and\ \citenamefont {Zeng}}]{Bern:2024vqs}%
  \BibitemOpen
  \bibfield  {author} {\bibinfo {author} {\bibfnamefont {Z.}~\bibnamefont {Bern}}, \bibinfo {author} {\bibfnamefont {E.}~\bibnamefont {Herrmann}}, \bibinfo {author} {\bibfnamefont {R.}~\bibnamefont {Roiban}}, \bibinfo {author} {\bibfnamefont {M.~S.}\ \bibnamefont {Ruf}}, \ and\ \bibinfo {author} {\bibfnamefont {M.}~\bibnamefont {Zeng}},\ }\href {\doibase 10.1007/JHEP06(2025)115} {\bibfield  {journal} {\bibinfo  {journal} {JHEP}\ }\textbf {\bibinfo {volume} {06}},\ \bibinfo {pages} {115} (\bibinfo {year} {2025}{\natexlab{b}})},\ \Eprint {http://arxiv.org/abs/2408.06686} {arXiv:2408.06686 [hep-th]} \BibitemShut {NoStop}%
\bibitem [{\citenamefont {Parra-Martinez}\ \emph {et~al.}(2020)\citenamefont {Parra-Martinez}, \citenamefont {Ruf},\ and\ \citenamefont {Zeng}}]{Parra-Martinez:2020dzs}%
  \BibitemOpen
  \bibfield  {author} {\bibinfo {author} {\bibfnamefont {J.}~\bibnamefont {Parra-Martinez}}, \bibinfo {author} {\bibfnamefont {M.~S.}\ \bibnamefont {Ruf}}, \ and\ \bibinfo {author} {\bibfnamefont {M.}~\bibnamefont {Zeng}},\ }\href {\doibase 10.1007/JHEP11(2020)023} {\bibfield  {journal} {\bibinfo  {journal} {JHEP}\ }\textbf {\bibinfo {volume} {11}},\ \bibinfo {pages} {023} (\bibinfo {year} {2020})},\ \Eprint {http://arxiv.org/abs/2005.04236} {arXiv:2005.04236 [hep-th]} \BibitemShut {NoStop}%
\bibitem [{\citenamefont {Saotome}\ and\ \citenamefont {Akhoury}(2013)}]{Saotome:2012vy}%
  \BibitemOpen
  \bibfield  {author} {\bibinfo {author} {\bibfnamefont {R.}~\bibnamefont {Saotome}}\ and\ \bibinfo {author} {\bibfnamefont {R.}~\bibnamefont {Akhoury}},\ }\href {\doibase 10.1007/JHEP01(2013)123} {\bibfield  {journal} {\bibinfo  {journal} {JHEP}\ }\textbf {\bibinfo {volume} {01}},\ \bibinfo {pages} {123} (\bibinfo {year} {2013})},\ \Eprint {http://arxiv.org/abs/1210.8111} {arXiv:1210.8111 [hep-th]} \BibitemShut {NoStop}%
\bibitem [{\citenamefont {Akhoury}\ \emph {et~al.}(2021)\citenamefont {Akhoury}, \citenamefont {Saotome},\ and\ \citenamefont {Sterman}}]{Akhoury:2013yua}%
  \BibitemOpen
  \bibfield  {author} {\bibinfo {author} {\bibfnamefont {R.}~\bibnamefont {Akhoury}}, \bibinfo {author} {\bibfnamefont {R.}~\bibnamefont {Saotome}}, \ and\ \bibinfo {author} {\bibfnamefont {G.}~\bibnamefont {Sterman}},\ }\href {\doibase 10.1103/PhysRevD.103.064036} {\bibfield  {journal} {\bibinfo  {journal} {Phys. Rev. D}\ }\textbf {\bibinfo {volume} {103}},\ \bibinfo {pages} {064036} (\bibinfo {year} {2021})},\ \Eprint {http://arxiv.org/abs/1308.5204} {arXiv:1308.5204 [hep-th]} \BibitemShut {NoStop}%
\bibitem [{\citenamefont {Jakobsen}\ \emph {et~al.}(2022)\citenamefont {Jakobsen}, \citenamefont {Mogull}, \citenamefont {Plefka},\ and\ \citenamefont {Steinhoff}}]{Jakobsen:2021zvh}%
  \BibitemOpen
  \bibfield  {author} {\bibinfo {author} {\bibfnamefont {G.~U.}\ \bibnamefont {Jakobsen}}, \bibinfo {author} {\bibfnamefont {G.}~\bibnamefont {Mogull}}, \bibinfo {author} {\bibfnamefont {J.}~\bibnamefont {Plefka}}, \ and\ \bibinfo {author} {\bibfnamefont {J.}~\bibnamefont {Steinhoff}},\ }\href {\doibase 10.1007/JHEP01(2022)027} {\bibfield  {journal} {\bibinfo  {journal} {JHEP}\ }\textbf {\bibinfo {volume} {01}},\ \bibinfo {pages} {027} (\bibinfo {year} {2022})},\ \Eprint {http://arxiv.org/abs/2109.04465} {arXiv:2109.04465 [hep-th]} \BibitemShut {NoStop}%
\bibitem [{\citenamefont {Duhr}\ and\ \citenamefont {Dulat}(2019)}]{Duhr:2019tlz}%
  \BibitemOpen
  \bibfield  {author} {\bibinfo {author} {\bibfnamefont {C.}~\bibnamefont {Duhr}}\ and\ \bibinfo {author} {\bibfnamefont {F.}~\bibnamefont {Dulat}},\ }\href {\doibase 10.1007/JHEP08(2019)135} {\bibfield  {journal} {\bibinfo  {journal} {JHEP}\ }\textbf {\bibinfo {volume} {08}},\ \bibinfo {pages} {135} (\bibinfo {year} {2019})},\ \Eprint {http://arxiv.org/abs/1904.07279} {arXiv:1904.07279 [hep-th]} \BibitemShut {NoStop}%
\bibitem [{\citenamefont {Weinzierl}(2022)}]{Weinzierl:2022eaz}%
  \BibitemOpen
  \bibfield  {author} {\bibinfo {author} {\bibfnamefont {S.}~\bibnamefont {Weinzierl}},\ }\href {\doibase 10.1007/978-3-030-99558-4} {\emph {\bibinfo {title} {{Feynman Integrals. A Comprehensive Treatment for Students and Researchers}}}},\ UNITEXT for Physics\ (\bibinfo  {publisher} {Springer},\ \bibinfo {year} {2022})\ \Eprint {http://arxiv.org/abs/2201.03593} {arXiv:2201.03593 [hep-th]} \BibitemShut {NoStop}%
\bibitem [{\citenamefont {Bourjaily}\ \emph {et~al.}(2022)\citenamefont {Bourjaily} \emph {et~al.}}]{Bourjaily:2022bwx}%
  \BibitemOpen
  \bibfield  {author} {\bibinfo {author} {\bibfnamefont {J.~L.}\ \bibnamefont {Bourjaily}} \emph {et~al.},\ }in\ \href@noop {} {\emph {\bibinfo {booktitle} {{Snowmass 2021}}}}\ (\bibinfo {year} {2022})\ \Eprint {http://arxiv.org/abs/2203.07088} {arXiv:2203.07088 [hep-ph]} \BibitemShut {NoStop}%
\bibitem [{\citenamefont {Chen}(1977)}]{Chen:1977oja}%
  \BibitemOpen
  \bibfield  {author} {\bibinfo {author} {\bibfnamefont {K.-T.}\ \bibnamefont {Chen}},\ }\href {\doibase 10.1090/S0002-9904-1977-14320-6} {\bibfield  {journal} {\bibinfo  {journal} {Bull. Am. Math. Soc.}\ }\textbf {\bibinfo {volume} {83}},\ \bibinfo {pages} {831} (\bibinfo {year} {1977})}\BibitemShut {NoStop}%
\bibitem [{\citenamefont {Brammer}\ \emph {et~al.}(2025)\citenamefont {Brammer}, \citenamefont {Frellesvig}, \citenamefont {Morales},\ and\ \citenamefont {Wilhelm}}]{Brammer:2025rqo}%
  \BibitemOpen
  \bibfield  {author} {\bibinfo {author} {\bibfnamefont {D.}~\bibnamefont {Brammer}}, \bibinfo {author} {\bibfnamefont {H.}~\bibnamefont {Frellesvig}}, \bibinfo {author} {\bibfnamefont {R.}~\bibnamefont {Morales}}, \ and\ \bibinfo {author} {\bibfnamefont {M.}~\bibnamefont {Wilhelm}},\ }\href {\doibase 10.1007/JHEP10(2025)212} {\bibfield  {journal} {\bibinfo  {journal} {JHEP}\ }\textbf {\bibinfo {volume} {10}},\ \bibinfo {pages} {212} (\bibinfo {year} {2025})},\ \Eprint {http://arxiv.org/abs/2505.10274} {arXiv:2505.10274 [hep-th]} \BibitemShut {NoStop}%
\bibitem [{\citenamefont {Ruf}(2023)}]{MRAmplitudes2023}%
  \BibitemOpen
  \bibfield  {author} {\bibinfo {author} {\bibfnamefont {M.}~\bibnamefont {Ruf}},\ }\href {https://indico.cern.ch/event/1228963/contributions/5506364/} {\enquote {\bibinfo {title} {Towards gravitational scattering at the fifth order in $g$},}\ } (\bibinfo {year} {2023}),\ \bibinfo {note} {{Amplitudes 2023 Conference}}\BibitemShut {NoStop}%
\bibitem [{\citenamefont {Joyce}\ and\ \citenamefont {Delves}(2004)}]{GSJoyce_2004}%
  \BibitemOpen
  \bibfield  {author} {\bibinfo {author} {\bibfnamefont {G.~S.}\ \bibnamefont {Joyce}}\ and\ \bibinfo {author} {\bibfnamefont {R.~T.}\ \bibnamefont {Delves}},\ }\href {\doibase 10.1088/0305-4470/37/20/012} {\bibfield  {journal} {\bibinfo  {journal} {Journal of Physics A: Mathematical and General}\ }\textbf {\bibinfo {volume} {37}},\ \bibinfo {pages} {5417} (\bibinfo {year} {2004})}\BibitemShut {NoStop}%
\bibitem [{\citenamefont {Ronveaux}(1995)}]{Ronveaux:1995Heun}%
  \BibitemOpen
  \bibinfo {editor} {\bibfnamefont {A.}~\bibnamefont {Ronveaux}},\ ed.,\ \href@noop {} {\emph {\bibinfo {title} {Heun's Differential Equations}}}\ (\bibinfo  {publisher} {Oxford University Press},\ \bibinfo {address} {Oxford},\ \bibinfo {year} {1995})\BibitemShut {NoStop}%
\bibitem [{\citenamefont {Ablinger}\ \emph {et~al.}(2018)\citenamefont {Ablinger}, \citenamefont {Bl{\"u}mlein}, \citenamefont {De~Freitas}, \citenamefont {van Hoeij}, \citenamefont {Imamoglu}, \citenamefont {Raab}, \citenamefont {Radu},\ and\ \citenamefont {Schneider}}]{Ablinger:2017bjx}%
  \BibitemOpen
  \bibfield  {author} {\bibinfo {author} {\bibfnamefont {J.}~\bibnamefont {Ablinger}}, \bibinfo {author} {\bibfnamefont {J.}~\bibnamefont {Bl{\"u}mlein}}, \bibinfo {author} {\bibfnamefont {A.}~\bibnamefont {De~Freitas}}, \bibinfo {author} {\bibfnamefont {M.}~\bibnamefont {van Hoeij}}, \bibinfo {author} {\bibfnamefont {E.}~\bibnamefont {Imamoglu}}, \bibinfo {author} {\bibfnamefont {C.~G.}\ \bibnamefont {Raab}}, \bibinfo {author} {\bibfnamefont {C.~S.}\ \bibnamefont {Radu}}, \ and\ \bibinfo {author} {\bibfnamefont {C.}~\bibnamefont {Schneider}},\ }\href {\doibase 10.1063/1.4986417} {\bibfield  {journal} {\bibinfo  {journal} {J. Math. Phys.}\ }\textbf {\bibinfo {volume} {59}},\ \bibinfo {pages} {062305} (\bibinfo {year} {2018})},\ \Eprint {http://arxiv.org/abs/1706.01299} {arXiv:1706.01299 [hep-th]} \BibitemShut {NoStop}%
\bibitem [{\citenamefont {Baikov}\ \emph {et~al.}(2017)\citenamefont {Baikov}, \citenamefont {Chetyrkin},\ and\ \citenamefont {K{\"u}hn}}]{Baikov:2016tgj}%
  \BibitemOpen
  \bibfield  {author} {\bibinfo {author} {\bibfnamefont {P.~A.}\ \bibnamefont {Baikov}}, \bibinfo {author} {\bibfnamefont {K.~G.}\ \bibnamefont {Chetyrkin}}, \ and\ \bibinfo {author} {\bibfnamefont {J.~H.}\ \bibnamefont {K{\"u}hn}},\ }\href {\doibase 10.1103/PhysRevLett.118.082002} {\bibfield  {journal} {\bibinfo  {journal} {Phys. Rev. Lett.}\ }\textbf {\bibinfo {volume} {118}},\ \bibinfo {pages} {082002} (\bibinfo {year} {2017})},\ \Eprint {http://arxiv.org/abs/1606.08659} {arXiv:1606.08659 [hep-ph]} \BibitemShut {NoStop}%
\bibitem [{\citenamefont {Barack}\ \emph {et~al.}(2023)\citenamefont {Barack}, \citenamefont {Bern}, \citenamefont {Herrmann}, \citenamefont {Long}, \citenamefont {Parra-Martinez}, \citenamefont {Roiban}, \citenamefont {Ruf}, \citenamefont {Shen}, \citenamefont {Solon}, \citenamefont {Teng},\ and\ \citenamefont {Zeng}}]{Barack:2023oqp}%
  \BibitemOpen
  \bibfield  {author} {\bibinfo {author} {\bibfnamefont {L.}~\bibnamefont {Barack}}, \bibinfo {author} {\bibfnamefont {Z.}~\bibnamefont {Bern}}, \bibinfo {author} {\bibfnamefont {E.}~\bibnamefont {Herrmann}}, \bibinfo {author} {\bibfnamefont {O.}~\bibnamefont {Long}}, \bibinfo {author} {\bibfnamefont {J.}~\bibnamefont {Parra-Martinez}}, \bibinfo {author} {\bibfnamefont {R.}~\bibnamefont {Roiban}}, \bibinfo {author} {\bibfnamefont {M.~S.}\ \bibnamefont {Ruf}}, \bibinfo {author} {\bibfnamefont {C.-H.}\ \bibnamefont {Shen}}, \bibinfo {author} {\bibfnamefont {M.~P.}\ \bibnamefont {Solon}}, \bibinfo {author} {\bibfnamefont {F.}~\bibnamefont {Teng}}, \ and\ \bibinfo {author} {\bibfnamefont {M.}~\bibnamefont {Zeng}},\ }\href {\doibase 10.1103/PhysRevD.108.024025} {\bibfield  {journal} {\bibinfo  {journal} {Phys. Rev. D}\ }\textbf {\bibinfo {volume} {108}},\ \bibinfo {pages} {024025} (\bibinfo {year} {2023})},\ \Eprint {http://arxiv.org/abs/2304.09200} {arXiv:2304.09200 [hep-th]} \BibitemShut {NoStop}%
\bibitem [{\citenamefont {Ivanov}\ \emph {et~al.}(2024)\citenamefont {Ivanov}, \citenamefont {Li}, \citenamefont {Parra-Martinez},\ and\ \citenamefont {Zhou}}]{Ivanov:2024sds}%
  \BibitemOpen
  \bibfield  {author} {\bibinfo {author} {\bibfnamefont {M.~M.}\ \bibnamefont {Ivanov}}, \bibinfo {author} {\bibfnamefont {Y.-Z.}\ \bibnamefont {Li}}, \bibinfo {author} {\bibfnamefont {J.}~\bibnamefont {Parra-Martinez}}, \ and\ \bibinfo {author} {\bibfnamefont {Z.}~\bibnamefont {Zhou}},\ }\href {\doibase 10.1103/PhysRevLett.132.131401} {\bibfield  {journal} {\bibinfo  {journal} {Phys. Rev. Lett.}\ }\textbf {\bibinfo {volume} {132}},\ \bibinfo {pages} {131401} (\bibinfo {year} {2024})},\ \bibinfo {note} {[Erratum: Phys.Rev.Lett. 134, 159901 (2025)]},\ \Eprint {http://arxiv.org/abs/2401.08752} {arXiv:2401.08752 [hep-th]} \BibitemShut {NoStop}%
\bibitem [{\citenamefont {Caron-Huot}\ \emph {et~al.}(2025)\citenamefont {Caron-Huot}, \citenamefont {Correia}, \citenamefont {Isabella},\ and\ \citenamefont {Solon}}]{Caron-Huot:2025tlq}%
  \BibitemOpen
  \bibfield  {author} {\bibinfo {author} {\bibfnamefont {S.}~\bibnamefont {Caron-Huot}}, \bibinfo {author} {\bibfnamefont {M.}~\bibnamefont {Correia}}, \bibinfo {author} {\bibfnamefont {G.}~\bibnamefont {Isabella}}, \ and\ \bibinfo {author} {\bibfnamefont {M.}~\bibnamefont {Solon}},\ }\href {\doibase 10.1103/qd3c-nfz6} {\bibfield  {journal} {\bibinfo  {journal} {Phys. Rev. Lett.}\ }\textbf {\bibinfo {volume} {135}},\ \bibinfo {pages} {191601} (\bibinfo {year} {2025})},\ \Eprint {http://arxiv.org/abs/2503.13593} {arXiv:2503.13593 [hep-th]} \BibitemShut {NoStop}%
\bibitem [{\citenamefont {Grozin}\ \emph {et~al.}(2016)\citenamefont {Grozin}, \citenamefont {Henn}, \citenamefont {Korchemsky},\ and\ \citenamefont {Marquard}}]{Grozin:2015kna}%
  \BibitemOpen
  \bibfield  {author} {\bibinfo {author} {\bibfnamefont {A.}~\bibnamefont {Grozin}}, \bibinfo {author} {\bibfnamefont {J.~M.}\ \bibnamefont {Henn}}, \bibinfo {author} {\bibfnamefont {G.~P.}\ \bibnamefont {Korchemsky}}, \ and\ \bibinfo {author} {\bibfnamefont {P.}~\bibnamefont {Marquard}},\ }\href {\doibase 10.1007/JHEP01(2016)140} {\bibfield  {journal} {\bibinfo  {journal} {JHEP}\ }\textbf {\bibinfo {volume} {01}},\ \bibinfo {pages} {140} (\bibinfo {year} {2016})},\ \Eprint {http://arxiv.org/abs/1510.07803} {arXiv:1510.07803 [hep-ph]} \BibitemShut {NoStop}%
\bibitem [{\citenamefont {Herrmann}\ \emph {et~al.}(2021{\natexlab{b}})\citenamefont {Herrmann}, \citenamefont {Parra-Martinez}, \citenamefont {Ruf},\ and\ \citenamefont {Zeng}}]{Herrmann:2021lqe}%
  \BibitemOpen
  \bibfield  {author} {\bibinfo {author} {\bibfnamefont {E.}~\bibnamefont {Herrmann}}, \bibinfo {author} {\bibfnamefont {J.}~\bibnamefont {Parra-Martinez}}, \bibinfo {author} {\bibfnamefont {M.~S.}\ \bibnamefont {Ruf}}, \ and\ \bibinfo {author} {\bibfnamefont {M.}~\bibnamefont {Zeng}},\ }\href {\doibase 10.1103/PhysRevLett.126.201602} {\bibfield  {journal} {\bibinfo  {journal} {Phys. Rev. Lett.}\ }\textbf {\bibinfo {volume} {126}},\ \bibinfo {pages} {201602} (\bibinfo {year} {2021}{\natexlab{b}})},\ \Eprint {http://arxiv.org/abs/2101.07255} {arXiv:2101.07255 [hep-th]} \BibitemShut {NoStop}%
\bibitem [{\citenamefont {Frellesvig}\ \emph {et~al.}(2025)\citenamefont {Frellesvig}, \citenamefont {Morales}, \citenamefont {P{\"o}gel}, \citenamefont {Weinzierl},\ and\ \citenamefont {Wilhelm}}]{Frellesvig:2024rea}%
  \BibitemOpen
  \bibfield  {author} {\bibinfo {author} {\bibfnamefont {H.}~\bibnamefont {Frellesvig}}, \bibinfo {author} {\bibfnamefont {R.}~\bibnamefont {Morales}}, \bibinfo {author} {\bibfnamefont {S.}~\bibnamefont {P{\"o}gel}}, \bibinfo {author} {\bibfnamefont {S.}~\bibnamefont {Weinzierl}}, \ and\ \bibinfo {author} {\bibfnamefont {M.}~\bibnamefont {Wilhelm}},\ }\href {\doibase 10.1007/JHEP02(2025)209} {\bibfield  {journal} {\bibinfo  {journal} {JHEP}\ }\textbf {\bibinfo {volume} {02}},\ \bibinfo {pages} {209} (\bibinfo {year} {2025})},\ \Eprint {http://arxiv.org/abs/2412.12057} {arXiv:2412.12057 [hep-th]} \BibitemShut {NoStop}%
\bibitem [{\citenamefont {Smirnov}(2004)}]{Smirnov:2004ym}%
  \BibitemOpen
  \bibfield  {author} {\bibinfo {author} {\bibfnamefont {V.~A.}\ \bibnamefont {Smirnov}},\ }\href@noop {} {\bibfield  {journal} {\bibinfo  {journal} {Springer Tracts Mod. Phys.}\ }\textbf {\bibinfo {volume} {211}},\ \bibinfo {pages} {1} (\bibinfo {year} {2004})}\BibitemShut {NoStop}%
\bibitem [{\citenamefont {Remiddi}\ and\ \citenamefont {Vermaseren}(2000)}]{Remiddi:1999ew}%
  \BibitemOpen
  \bibfield  {author} {\bibinfo {author} {\bibfnamefont {E.}~\bibnamefont {Remiddi}}\ and\ \bibinfo {author} {\bibfnamefont {J.~A.~M.}\ \bibnamefont {Vermaseren}},\ }\href {\doibase 10.1142/S0217751X00000367} {\bibfield  {journal} {\bibinfo  {journal} {Int. J. Mod. Phys. A}\ }\textbf {\bibinfo {volume} {15}},\ \bibinfo {pages} {725} (\bibinfo {year} {2000})},\ \Eprint {http://arxiv.org/abs/hep-ph/9905237} {arXiv:hep-ph/9905237} \BibitemShut {NoStop}%
\end{thebibliography}%

\end{document}